\documentclass[final,3p,onecolumn]{elsarticle}

\usepackage{amssymb,epsfig,latexsym,times,bbm,lineno}
\usepackage{amsmath}
\usepackage{breqn}
\usepackage{color}
\usepackage{mathtools, cuted}

\usepackage[percent]{overpic}
\usepackage{color}
\usepackage{graphicx}
\usepackage[export]{adjustbox}
\usepackage{dcolumn}
\usepackage{blkarray}
\usepackage{url}
\usepackage[skip=0.333\baselineskip]{caption}
\usepackage{subcaption}

\begin{document}

\begin{frontmatter}

\title{Heterogeneous fragmentation of empty sites promotes cooperation \\ in phenotypically diverse populations with tag-mediated interactions}

\author[a,1]{Hui Zhang}
\ead{huizhang@nwpu.edu.cn}
\author[b,1]{Tarik Hadzibeganovic}
\ead{tarik.hadzibeganovic@gmail.com}
\author[c,d]{Xiao-Pu Han}
\ead{xp@hznu.edu.cn}

\address[a]{School of Mathematics and Statistics, Northwestern Polytechnical University, Xi'an, Shaanxi 710072, China}
\address[b]{Institute of Psychology, Faculty of Natural Sciences, University of Graz, Graz 8010, Austria}
\address[c]{Alibaba Research Center for Complexity Sciences, Hangzhou Normal University, Hangzhou 311121, China}
\address[d]{Institute of Information Economy and Alibaba Business School, Hangzhou Normal University, Hangzhou 311121, China}

\footnotetext[1]{These authors contributed equally and share first authorship.}

\begin{abstract} 

Habitat loss and fragmentation have often been viewed as major threats to species interaction and global biodiversity conservation. However, habitat degradation can also give rise to positive ecological and behavioral responses, challenging the notion that its consequences are entirely detrimental. While controlling for the degree of total habitat loss, we studied the influence of habitat fragmentation and phenotypic diversity on the evolution of tag-based cooperation in structured populations with multiple strategies. We developed a spatially explicit agent-based model with empty sites in which phenotypically diverse artificial decision makers engaged into pairwise Snowdrift-game interactions and imitated strategies of their opponent co-players. We systematically varied the number of phenotypic features in the population, the clustering degree of empty sites unsuitable for habitation, as well as the cost-to-benefit ratio $r$, and we measured the resulting equilibrium densities of conditional and unconditional strategies. Our Monte Carlo simulations revealed a complex interplay between the three investigated factors, such that higher phenotypic diversity in combination with lower $r$ and low to intermediate clustering degrees of empty sites markedly suppressed ethnocentric cooperation but simultaneously boosted unconditional, pure altruism. This dominance of unconditional cooperation was remarkably robust to variation in the initial conditions, suggesting that heterogeneous fragmentation of empty sites in moderately degraded habitats can function as a potent cooperation-promoting mechanism even in the presence of initially more favorable strategies. Our study showcases anti-fragility of cooperators in spatially fragmented but phenotypically diverse populations, as they were also able to benefit from harsh environmental conditions emerging in sparsely connected habitat remnants.

\end{abstract}


\end{frontmatter}



\section{Introduction}
\label{}

Climate change, dryland expansion, and the associated habitat degradation and loss represent some of the major threats to biodiversity conservation and species interaction on our planet \cite{pimmraven}. Habitat destruction increases the overall competition for the ever scarcer resources \cite{lionvanbaalen}, whereas its further fragmentation can considerably disrupt metapopulation dynamics \cite{cheptou,fahrig,sardenyassole,fahrig2}, leading to a critically reduced gene flow and a substantial decline of genetic diversity. 

Although detrimental consequences of habitat loss for biological diversity have been widely documented \cite{fahrig,fahrig2,parkermacnally,limafilho2021}, it remains rather unclear if the effects of habitat fragmentation \textit{per se} are exclusively negative \cite{fahrig2,ZhangBearupetal2023,NatEcolEvol2024}. For example, besides the increased scarcity of resources and the emerging asymmetries in inter-species contacts across fragmented habitats, spatial fragmentation of habitable sites can also promote species coexistence \cite{NatEcolEvol2024}, reduce competition and the overall extinction risk \cite{fahrig2}, 
or elevate prosocial behavior in the population of isolated individuals \cite{zhanghuihan}.  

Interestingly, habitat loss and fragmentation have often been reported to interact, i.e. the actual effects of fragmentation or habitat subdivision on biodiversity seem to depend on the underlying degree of habitat loss \cite{fahrig,fahrig2,NatEcolEvol2024}, even though some studies have yielded rather mixed findings regarding these claimed dependencies \cite{fahrig2,parkermacnally,taoetal2024}. In addition, apart from research on ecological responses to fragmented habitats \cite{fahrig2}, systematic investigations of \textit{behavioral} responses to habitat loss and fragmentation, including the studies of cooperative behavior in destructed and fragmented habitats, are relatively recent and still lacking \cite{alizon,sekiguchi,yangphysa,yangpla,zhangetalempty}.

Ever since the seminal work of Hamilton \cite{hamiltonfish}, and then two decades later by Nowak and May \cite{nowakmay}, clarifying the roles that spatial structure can play in the evolution of cooperation has been one of the most intensively investigated topics in evolutionary game theory \cite{tarnita,ohtsuki,randnowak}. 
In spite of a tremendous amount of knowledge that has been gathered over the past few decades on the dynamics of evolutionary games in structured populations  \cite{tarnita,ohtsuki,fiverules,nandadurret,masudahetero,tomassini,zhangmathbiol,vanveelen,huilattice,allenetal}, our understanding of the actual effects that interaction structure imparts on cooperative behavior still remains elusive \cite{gracialazaro,cuestasanchez,grujic}, and the topic continues to attract an ever increasing attention across disciplines \cite{mcavoy,shixiao,sinharoy,svoboda2024,linli2025}.

In tag-based cooperation \cite{holland,riolo,hamaxel}, which has meanwhile emerged as a distinct area in evolutionary game theory research \cite{traulsenandnowak,antal,laird1,garciaveelen,hadzibeganovic0,hadzibeganovic1,hartshorn,hadzibeganovic2,hadzibeganovic3,ramazi,morsky,Jensen2019pre,hadzibeganovic4,hadzibeganovic5,dhakal,jeongetal,HeLiDu2025,HeChengLu2025,efferson,masufu}, spatial interaction structure is not a necessary prerequisite for the emergence of prosocial behavior, because tag-based generosity can also evolve and thrive through interactions occurring merely in the virtual space of phenotypic features \cite{antal,hadzibeganovic4,efferson}. Tag-mediated cooperation can thus evolve in the absence of any explicit population structure, which can 
otherwise be highly detrimental to the emergence of any cooperative venture \cite{ohtsuki}. However, even though it is not essentially required for the evolution of tag-based cooperation, spatial interaction structure can undoubtedly enhance it \cite{traulsenandnowak,hadzibeganovic4}, especially when 
it is combined with other potent cooperation-inducing mechanisms \cite{hadzibeganovic3,hadzibeganovic5}. 

Tag-based cooperation is a type of benevolent behavior directed nepotistically towards sufficiently similiar or identical others, i.e. tag-mates who share discernible phenotypic traits. Although tag-based generosity has meanwhile been acknowledged as a widespread phenomenon in nature \cite{masufu}, its adequate evolutionary explanation is still missing \cite{randnowak}.

Notably, in both tagless and tag-based cooperation models, the underlying population structure has usually been oversimplified by employing mostly fully inhabitable, regular square-lattice-based spatial configurations \cite{hamaxel,hartshorn,masufu,santospalattice}. Another yet largely neglected issue pertaining to methodological aspects of evolutionary game theoretic studies (both tagless and tag-based), is the often observed strong bias towards reporting only positive effects of network structure on cooperation. Clearly, in evolutionary game theory we want to identify the underlying mechanisms and combinations thereof that consistently facilitate the evolution of prosocial behavior. However, in a number of cases, as in populations with tag-mediated interactions and conditional strategies, causing multiple possible equilibria \cite{ramazi,hadzibeganovic5,jeongetal} or oscillatory out-of-equilibrium phenomena \cite{traulsenandnowak}, it is of particular interest to identify conditions under which discriminating but potent strategies such as ingroup-biased ethnocentrism can be efficiently controlled, while simultaneously promoting socially more desirable strategies such as unconditional cooperation. 

One way to effectively circumvent oversimplification in evolutionary games on graphs and networks and to address more realistic structural effects on cooperation, without necessarily employing excessively complex networked systems, is to investigate heterogeneous spatial configurations \cite{masudahetero,gracialazaro,galliard} devised via minimal but non-trivial modifications of the basic lattice structure. For example, in the case of environmental heterogeneities that emerge under habitat loss and fragmentation \cite{hiebeler1}, one can introduce remarkable levels of spatial diversity by simply allowing for a fraction of network sites to be empty \cite{lionvanbaalen}, either temporarily (where they can later be populated) \cite{alizon} or permanently (where they can represent unsuitable sites, unfit for habitation) \cite{hiebeler1}. 

Interestingly, previous game-theoretic studies considering these spatial heterogeneities \cite{alizon,sekiguchi,yangphysa,yangpla} have revelead that intermediate amounts of empty sites \cite{alizon} or moderately destructed habitats \cite{yangphysa,yangpla} can indeed promote indiscriminate cooperation, but they cannot decrease the degree of spiteful behavior \cite{sekiguchi}. However, these earlier studies did not address the question of how precisely this moderate amount of habitat loss needs to be distributed across the landscape in order to optimally promote not only cooperation, but also the evolution of other unconditional and conditional competing strategies. They thus did not consider the presence of correlation patterns in structured heterogeneities of empty sites, and were mostly conducted in the context of either prisoner's dilemma (PD) \cite{yangphysa} or public goods (PG) games \cite{yangpla}. In addition, the study in Ref. \cite{alizon} actually assumed that empty sites enable the emergence of more 'free space' in the environment so that individuals with higher fecundity can still use it for their produced offspring, neglecting thereby the existence of detrimental effects of permanently uninhabitable empty sites that are widespread in natural habitats. 

The study of Zhang and colleagues \cite{zhangetalempty} was to our knowledge the first one to explicitly investigate structured environmental heterogeneities with empty sites and their influence on the evolution of strategies in both PD and Snowdrift (SD) games, finding remarkable benefits of autocorrelated empty sites for cooperation in PD games, but variable outcomes in the SD game depending on the benefit of cooperation $b$. For example, in a model with SD game interactions and heterogeneously structured emtpy sites \cite{zhangetalempty}, cooperation evolved at a larger clustering degree of emtpy sites but only if temptation to defect $b$ was sufficiently high, whereas at intermediate and lower values of $b$ cooperation was prevented from fixation.  

Nevertheless, it needs to be stressed here that the study of Zhang and colleagues \cite{zhangetalempty}, as well as other investigations in this research domain \cite{alizon,sekiguchi,yangphysa,yangpla}, considered only tagless models of cooperation with mostly two pure strategies (or generalizations thereof), therefore including only phenotypically homogeneous populations. However, to the best of our knowledge, the effects of fragmented landscapes with heterogeneously distributed empty sites on cooperation have never been investigated in phenotypically diverse populations with multiple competing strategies. Moreover, besides earlier theoretical results, the outcomes of many empirical studies in this domain also remain largely inconclusive, as landscape fragmentation and the associated dissolution of habitats were not measured while controlling for habitat loss \cite{fahrig,fahrig2}, and hence the resulting measures of the two variables were often confounded, rendering the interpretation of the reported findings difficult. 

In other words, it has not been clarified yet how the variation in empty site distribution under constant amount of unsuitable sites can affect the evolution of multiple conditional and unconditional strategies in tag-based cooperation models, and under precisely what ecological conditions can pure altruism and strategies other than ingroup-biased ethnocentism thrive and prevail in phenotypically heterogeneous populations. Our present investigation is therefore the first one to systematically address the effects of spatially correlated empty sites on the evolution of multiple strategies in structured populations with tag-mediated interactions. 

Addressing the role of uninhabitable sites in the evolution of cooperation is particularly relevant due to a central shortcoming of previous studies considering largely idealistic ecological conditions under which empty sites did not exist and were not allowed to emerge in the course of the system evolution, i.e. even in the presence of birth-death dynamics, individual agents in many earlier models were typically instantly replaced by other agents and the system was therefore always fully saturated throughout the simulation. However, in tag-based cooperation models that followed the tradition of Hammond and Axelrod \cite{hamaxel}, the emergence of empty sites was usually explicitly enabled, but their effects on cooperation were never really systematically investigated. 

Acknowledging the existence of empty sites is also of general importance for any evolutionary model of cooperation, because the results of competition for reproductive shares are usually heavily affected by the occurring limitations in the amount of system sites that are effectively suitable for reproduction \cite{nonacs}. By ignoring these aspects, many previous models were not capable of addressing the roles that variation in reproduction site 
suitability could have played in the evolution of cooperation, e.g. in the presence of partially destructed and heterogeneously fragmented habitat sites unsuitable for reproduction (for a related discussion, see Ref. \cite{hadzibeganovic5}). 

Although a few related models have investigated coordination games in phenotypically diverse populations \cite{janssonjtb,barreira}, including Snowdrift games with different types of tags \cite{greenwood}, only one study has thus far {\it systematically} addressed the evolutionary dynamics of tag-based cooperation in the context of both aspatial and spatial Snowdrift games \cite{laird1}. In the Snowdrift game, the best option is always to adopt the strategy that is different from that of the opponent, which can often serve as a mechanism leading to stable coexistence of strategies in well-mixed populations \cite{pdsd}. 

However, in tagless but spatial Snowdrift games, this same mechanism in combination with the limited amount of local interactions can also produce the opposite effect, by reducing or even fully suppressing cooperative behavior to vanishingly low levels \cite{haudo2004}. On the other hand, in tag-based spatial Snowdrift games \cite{laird1}, a wide variety of outcomes is possible, including the dominance of either conditional cooperators or unconditional defectors, as well as the different types of coexistence of the two named strategies. Thus, unlike the {\it aspatial} Snowdrift game, in which resulting outcomes are 
somewhat straightforward, the conditions required for the promotion of cooperation in {\it spatial} Snowdrift games are generally less well understood \cite{pdsd}, and as such, they have never been subject to investigation in populations with tag-based interactions evolving on spatial structures with heterogeneously fragmented empty sites. 

Motivated by these gaps in previous research, we used a specific algorithm for generating spatially heterogeneous systems with empty sites that allowed us to precisely define their fragmention statistics and to investigate for the very first time their influence on the evolution of strategies in structured populations playing Snowdrift games with tag-mediated interactions (while simultaneously controlling for the overall amount of habitat loss in the model). Using this spatially explicit model, our primary focus was to find out if heterogeneous fragmentation of unsuitable sites in moderately degraded habitats can function as a mechanism for the promotion of indiscriminate cooperation and control of ingroup-biased or other conditional strategies in phenotypically diverse populations. 

In typical natural scenarios, factors influencing cooperative behavior almost always operate in
combination, i.e. jointly with other potent mechanisms \cite{randnowak}. Nevertheless, most previous studies of the evolution of cooperation were limited to the assessment of individual mechanisms in isolation, in spite of the fact that their combined influences were often found to 
generate intriguingly novel evolutionary outcomes \cite{randnowak,vanveelen,garciaveelen,hadzibeganovic3,hadzibeganovic5}. Notwithstanding valuable insights gleaned over the recent years about the joint effects of two distinct mechanisms on the evolution of 
cooperation (e.g. \cite{tanimotofactorial,taniki}), our understanding of the complex interaction effects emerging between three or more combined factors in tag-based evolutionary games still remains incomplete (see e.g. \cite{hadzibeganovic3}). In our current paper, we therefore aim at contributing to this expanding line of research by studying primarily how the density of unconditional cooperators (i.e. altruists) may depend upon the 
interplay between the clustering degree of empty sites and other relevant factors such as phenotypic diversity of the population and cost-to-benefit ratio. 

More specifically, we expected that under reasonably low cost-to-benefit ratios $r$, phenotypic diversity ($\lambda$) could moderate the relationship between the clustering degree of empty sites ($q_{0|0}$) and the density of altruists ($\rho_C$) such that the dependence of $\rho_C$ on $q_{0|0}$ should be stronger when $\lambda$ is high than when $\lambda$ is low. In contrast, at considerably higher $r$, larger phenotypic diversity $\lambda$ was not assumed to mitigate the expected negative effect of elevated $q_{0|0}$ on $\rho_C$. We thus hypothesized that the moderating effect of $\lambda$ on the association between $q_{0|0}$ and $\rho_C$ should vary as a function of $r$, suggesting thus a non-trivial complex interplay between the studied factors.

In addition, we aimed at a better understanding of the phenomenon of strategy coexistence in evolutionary games occuring in heterogeneously fragmented populations with tag-mediated interactions. Given that only a limited number of studies observed species coexistence in fragmented habitats \cite{NatEcolEvol2024} and strategy coexistence phenomena in tag-based cooperation models (e.g. \cite{laird1,hadzibeganovic2,hadzibeganovic5}), we were particularly interested in knowing for which strategies and precisely under what conditions is this coexistence feasible in our present model with degraded habitat structure and inhomogeneously distributed empty sites. 

From the analytical point of view, predicting the effects that (heterogeneous) space imparts on the coexistence of strategies is rather difficult \cite{szabofath2007}, which is why we need to rely on systematic computer-based simulations to understand the spatial evolutionary dynamics of multiple competing strategies \cite{laird1,adamiabm}. However, to fully appreciate the impact that fragmented interaction structure may have on the evolution of strategies in our spatial model, we first investigated analytically a minimal, aspatial model of tag-based cooperation. The outcomes of this minimal mathematical model then served as a baseline against which we compared and discussed our findings obtained from the corresponding computational model with an explicitly implemented spatial interaction structure with empty sites.

\section{The basic model}
\subsection{Snowdrift game}
Apart from the PD game \cite{axelhamil}, the SD game, also known as the Hawk-Dove or the Chicken game \cite{maynard}, is another often employed, alternative game-theoretic paradigm in the study of evolutionary dynamics of cooperation \cite{doebelihau}. In a two-player SD game with two fixed strategies, cooperation ($C$) and defection ($D$), the payoff matrix can be defined as follows

\begin{equation}
\begin{array}{cc}
&C~~~~D\\
\begin{array}{c}
    C\\
    D\\
  \end{array}& \left(
  \begin{array}{cc}
    R &~ S\\
    T &~ P\\
 \end{array}
\right).\\
\end{array}
\end{equation}

\noindent Here, if both players cooperate, their mutual cooperation will be acknowledged with the reward $R$. If they both
defect, which yields the worst possible outcome, they will be confronted with the punishment $P$. If one player, however, refuses to cooperate while the
other one decides to provide help, the defector then receives the temptation payoff $T$, whereas the cooperating player obtains the sucker's payoff $S$. The corresponding payoff ranking in the Snowdrift game is then given by $T>R>S>P$. 

In terms of the associated benefits $b$ and costs $c$, the four parameters can be further rewritten as $R= b-c/2$, $S=b-c$, $T=b$, and $P=0$, and the cost-to-benefit ratio is defined as $r=c/(2b-c)$. Substituting these expressions into the above defined payoff matrix (1) gives us:  

\begin{equation}
\begin{array}{cc}
&C~~\,~~D\\
\begin{array}{c}
    C\\
    D\\
  \end{array}& \left(
  \begin{array}{cc}
    b-\frac{c}{2} &~ b-c\\
    b &~ 0\\
 \end{array}
\right).\\
\end{array}
\end{equation}

Notably, in contrast to the PD game, in the tagless SD game played in well-mixed, large populations, cooperators and defectors can both stably coexist in evolutionary time whereby the equilibrium frequency
of cooperating agents is given by $X^{*}_{C} = (P - S) / (R-S-T+P)$, and the equilibrium frequency of defectors is $X^{*}_{D} = 1 - X^{*}_{C}$ \cite{haudo2004,hofsig98,nowak2006}.

\subsection{Two-tags model}
In a typical tag-based cooperation model \cite{hamaxel}, tags are usually defined as phenotypic features or markers (e.g. as colors) that are easily discernible by all competing players. Independent of tags, in the standard model of tag-based cooperation \cite{hamaxel} individuals can employ one of the four possible strategies: Unconditional cooperation or pure altruism ($C$), conditional 'intragroup' cooperation or ethnocentrism ($I$), conditional 'extragroup' cooperation or cosmopolitanism ($E$), and unconditional defection or egoism ($D$). 

Tag color and strategy are randomly assigned but they remain fixed throughout an individual's lifetime. An individual with the strategy $C$ always cooperates with everyone else in the population irrespective of their displayed tags. An individual with strategy $D$, however, continuously refuses to cooperate with all opponent co-players irrespective of their displayed tags. 

Tag-based strategies, $I$ and $E$, represent the two specific types of tag-mediated conditional cooperation. The strategy $I$ represents intra-tag cooperation (also known as 'ethnocentrism') whereby individuals cooperate nepotistically, i.e. only with identical tag-mates who share the identical tag color; otherwise, they always defect. The strategy $E$ is extra-tag cooperation which occurs only among opponents with different tag colors; otherwise, whenever two tag-mates with an identical tag color and strategy $E$ meet each other, they will always defect in the model. 

The resulting $8 \times 8$ payoff matrix, with two distinct tags and four strategies in the Snowdrift game \cite{laird1} is given by $\mathbf{A}=[A_{i,j}]=$
\begin{equation}
\hspace*{-3em}
\begin{array}{cc}
&C_1~~~~~~I_1~~~~~~E_1~~~~~~D_1~~~~~~C_2~~~~~~~I_2~~~~~~~E_2~~~~~~D_2\\ \\
\begin{array}{c}
    C_1\\ \\
    I_1\\ \\
   E_1\\ \\
D_1\\ \\
C_2\\ \\
I_2\\ \\
E_2\\ \\
D_2\\
  \end{array}& \left(
  \begin{array}{cccccccc}
    	R &~~~~ R &~~~~S &~~~~S &~~~~ R &~~~~S  &~~~~R  &~~~~S \\ \\
    R &~~~~ R &~~~~S &~~~~S  &~~~~ T  &~~~~P  &~~~~T  &~~~~ P \\ \\
   T &~~~~ T &~~~~P &~~~~P &~~~~R  &~~~~S &~~~~R  &~~~~S \\ \\
   T &~~~~ T &~~~~P &~~~~P &~~~~ T  &~~~~P &~~~~T &~~~~ P \\ \\
R &~~~~ S &~~~~R &~~~~S &~~~~R &~~~~R  &~~~~S  &~~~~ S \\ \\
 T &~~~~ P &~~~~T &~~~~P &~~~~ R &~~~~R &~~~~S  &~~~~ S \\ \\
R &~~~~S &~~~~R &~~~~S  &~~~~T &~~~~T  &~~~~P &~~~~ P \\ \\
T &~~~~ P &~~~~T &~~~~P  &~~~~T  &~~~~T  &~~~~P  &~~~~P \\
\end{array}
\right),\\
\end{array}
\end{equation}
\medskip
\smallskip

\noindent where $A_{ij}$ is the payoff of an individual player with the tag (subscript number) and strategy (letter) of row $i$ when interacting with an opponent co-player of the tag and strategy of column $j$. 

\subsection{Imitation rule}
A number of learning mechanisms have been introduced thus far in evolutionary game theoretic models for the adoption of new strategies. One frequently employed example is the so-called imitation rule \cite{traulsen2006}. Here, a randomly selected focal player compares its current payoff against the payoff of the randomly chosen opponent co-player, and the focal player then switches its strategy to that of the opponent with probability given by the Fermi-like function \cite{szabofermi,traulsen371}:
\begin{equation}
p=\frac{1}{1+e^{-\beta (\pi_{c}-\pi_{f})}}
\end{equation}
\noindent where $\beta$ stands for the intensity of selection, and $\pi_{f}$ and $\pi_{c}$ are the payoffs of the two randomly selected co-players. In the limit of $\beta \ll 1$, selection is weak and the underlying dynamics correspond to the frequency-dependent Moran process \cite{traulsen371}.

\section{Aspatial model}
\subsection{Minimal model for tag-based cooperation}
To fully understand the effects that spatial structure with empty sites may have on the evolution of strategies in a model with tag-mediated interactions, we first investigate a corresponding aspatial model of tag-based cooperation, i.e. a model without any explicitly defined interaction structure. Such a minimal, aspatial tag-based cooperation model, with the finite number $K$ of tags, has been introduced previously by Traulsen \cite{traulsen371}. 

Here, we first review the associated finite population framework, which was employed in Ref. \cite{traulsen371} for the systems containing $K$ tags with two and three strategies (and therefore with $2K$ and $3K$ different types of individuals). We then employ this framework to generalize the minimal model to the system with $K$ tags and four strategies, and therefore, with $4K$ individual types.

More generally, for the two strategies $P$ and $Q$, we let the payoff matrix be defined as 
\begin{equation}
\begin{array}{cc}
&P~~~~Q\\
\begin{array}{c}
    P\\
    Q\\
  \end{array}& \left(
  \begin{array}{cc}
    \widetilde a &~ \widetilde b\\
    \widetilde c &~ \widetilde d\\
 \end{array}
\right).\\
\end{array}
\end{equation}
With a very low probability $\mu \ll 1$, an individual may consider trying out a completely novel strategy, which may, however, not yet be available in the considered population \cite{traulsen371}. However, even though errors of this kind are rather rare, an occassional error may emerge and the novel strategy 
can then invade and ultimately dominate the population. We then consider a scenario with $l$ individuals playing $P$ strategy and $N-l$ individuals playing $Q$ strategy. The payoff of the $P$-strategists is then given by \cite{traulsen371}
\begin{equation}
\pi^{P}(l)=\widetilde a\frac{l-1}{N-1}+\widetilde b\frac{N-l}{N-1}
\end{equation}
\noindent and correspondingly for the $Q$-strategists we have 
\begin{equation}
\pi^{Q}(l)=\widetilde c\frac{l}{N-1}+\widetilde d\frac{N-l-1}{N-1}.
\end{equation}
A number of $P$ individuals can then change from $l$ to $l\pm 1$ with the probability \cite{traulsen371}
\begin{equation}
T^{\pm}(l)=\frac{l}{N}\frac{N-l}{N}\frac{1}{1+e^{\pm\beta(\pi^{P}-\pi^{Q})}},
\end{equation}
\noindent and a single type $P$-strategist can take over the population of type $Q$-strategists with the probability \cite{traulsen371}
\begin{equation}
\phi_{Q \rightarrow P}=\left[1+\sum^{N-1}_{m=1}\prod\limits_{l=1}^{m}\frac{T^{-}(l)}{T^{+}(l)}\right]^{-1}.
\end{equation}

Focusing on the weak selection scenario $\beta  \ll 1$, we then finally have that \cite{traulsen371}
\begin{equation}
\phi_{Q \rightarrow P}=\frac{1}{N}+\frac{\beta}{6}\left(\widetilde a+2\widetilde b-\widetilde c-2\widetilde d-\frac{2\widetilde a+\widetilde b+\widetilde c-4\widetilde d}{N}\right),
\end{equation}
\noindent which is essentially equivalent to the results reported in earlier studies for the frequency-dependent Moran process \cite{traulsen371}.

\subsection{Analytical results for SD games with $4K$ types of individuals}
\medskip

Adhering to this analytical method, we now provide the analysis for our four-strategy game with the payoff matrix (3) and with $4K$ types of individuals. In analogy to Ref. \cite{traulsen371}, we first take the transition probability from strategy $C$ to strategy $I$ as an example. We thus first compute the rate at which the population transitions from unconditional cooperation $C$ to intra-group conditional cooperation $I$. 

Thus, we consider a scenario where a $C_{1}$ population is taken over by an $I_{1}$ population, in which a single mutant can fixate with probability $\frac{1}{N}-\frac{\beta}{6}(\frac{3b-\frac{3c}{2}}{N})$ \cite{traulsen371}. Such mutants are typically generated at the rate $\mu/4K$, and the corresponding transition rate is thus given by $\phi_{C_{1} \rightarrow I_{1}}=\mu/(4K) ( \frac{1}{N}-\frac{\beta}{6} ( \frac{3b-\frac{3c}{2}}{N}))$. 

Next, the population of the type $C_{1}$ can also be invaded by the $I_{2}$ population of ingroup-biased individuals, which is advantageous in this case. The rate at which individuals of this particular type appear is given by $\mu(K-1)/(4K)$, and the probability of $I_{2}$ individuals taking over the population of $C_{1}$ strategists with a different tag is therefore
\begin{equation}
\phi_{C_{1} \rightarrow I_{2}}=\frac{1}{N}+\frac{\beta}{6}\left(\frac{3c}{2}-\frac{3b-\frac{3c}{2}}{N}\right).
\end{equation}
The joint rate of the processes leading from $C$ to $I$, $\rho_{C \rightarrow I}=\frac{\mu}{4K}\phi_{C_{1} \rightarrow I_{1}}+\frac{\mu}{4}\phi_{C_{1} \rightarrow I_{2}}$, is then given by
\begin{equation}
\rho_{C \rightarrow I}=\frac{\mu}{4K}\left(\frac{1}{N}-\frac{\beta}{6}\left(\frac{3b-\frac{3c}{2}}{N}\right)\right)+\mu\frac{K-1}{4K}\left(\frac{1}{N}+\frac{\beta}{6}\left(\frac{3c}{2}-\frac{3b-\frac{3c}{2}}{N}\right)\right).
\end{equation}
Similarly, other transition probabilities can be calculated with this same method. For example, the transition probability from $D$ to $I$, where the population of unconditional defectors can be invaded by conditional, ethnocentric cooperators, is given by 
\begin{equation}
\rho_{D \rightarrow I}=\frac{\mu}{4K}\left(\frac{1}{N}+\frac{\beta}{6}\left(2b-\frac{5c}{2}-\frac{4b-2c}{N}\right)\right)+\mu\frac{K-1}{4K}\left(\frac{1}{N}
+\frac{\beta}{6}\left(b-\frac{c}{2}-\frac{2b-c}{N}\right)\right).
\end{equation}

The remaining transition probabilities between individual strategies in our aspatial model are presented in the Appendix A. Finally, considering these obtained transition rates, we can further establish a transition matrix for the four competing strategies \cite{traulsen371,hauertetal2007}. The largest eigenvector gives the stationary distribution, and it reads

\begin{equation}
\left[
\begin{matrix}
p_C^{'}\\\\
p_I^{'}\\\\
p_E^{'}\\\\
p_D^{'}\\\\
\end{matrix}
\right]=\left[
\begin{matrix}
\frac{1}{4}\\\\
\frac{1}{4}\\\\
\frac{1}{4}\\\\
\frac{1}{4}
\end{matrix}
\right]+\frac{N\beta}{32}\left[
\begin{matrix}
\frac{c-2b}{N}+2b-3c\\\\
\frac{c-2b}{N}+2b+c-\frac{4c}{K}\\\\
\frac{2b-c}{N}-2b-c+\frac{4c}{K}\\\\
\frac{2b-c}{N}-2b+3c
\end{matrix}
\right]
+ \left[
\begin{matrix}
o(\beta)^2\\\\
o(\beta)^2\\\\
o(\beta)^2\\\\
o(\beta)^2
\end{matrix}
\right].
\end{equation}

For very large populations $N \gg 1$ and under the constraint of weak selection pressure $\beta \ll 1$ (and $\beta N \ll 1$), we then obtain
\begin{equation}
\left[
\begin{matrix}
p_C^{'}\\\\
p_I^{'}\\\\
p_E^{'}\\\\
p_D^{'}\\\\
\end{matrix}
\right]=\left[
\begin{matrix}
\frac{1}{4}\\\\
\frac{1}{4}\\\\
\frac{1}{4}\\\\
\frac{1}{4}
\end{matrix}
\right]+\frac{N\beta}{32}\left[
\begin{matrix}
2b-3c\\\\
2b+c-\frac{4c}{K}\\\\
-2b-c+\frac{4c}{K}\\\\
-2b+3c
\end{matrix}
\right].
\end{equation}

In particular, for the minimal $K=2$ tags case, the equation (15) can be further simplified to
\begin{equation}
\left[
\begin{matrix}
p_C^{'}\\\\
p_I^{'}\\\\
p_E^{'}\\\\
p_D^{'}
\end{matrix}
\right]=\left[
\begin{matrix}
\frac{1}{4}\\\\
\frac{1}{4}\\\\
\frac{1}{4}\\\\
\frac{1}{4}
\end{matrix}
\right]+\frac{N\beta}{32}\left[
\begin{matrix}
2b-3c\\\\
2b-c\\\\
-2b+c\\\\
-2b+3c
\end{matrix}
\right].
\end{equation}
We can therefore conclude that, whenever the condition $1<\frac{b}{c}<\frac{3}{2}$ is satisfied, the ranking of the frequencies of the four strategies in our aspatial model will be $p_E<p_C<p_D<p_I$, and when $\frac{b}{c} > \frac{3}{2}$, the ranking of strategies will be $p_E<p_D<p_C<p_I$. It is easy to find that the abundance of intragroup cooperators $I$ is always the largest in this aspatial model, whereas the frequency of extragroup cooperators $E$ is always the lowest among the four competing strategies. The relative ranking of the frequency of unconditional cooperators $C$ and unconditional defectors $D$ depends thus on the model parameters. We also found that in this aspatial model the abundance of $I$ strategists is always larger than 1/4 of the population, whereas the abundance of extra-tag cooperators $E$ is always less than the neutral case of 1/4.  

\section{Spatial model}
\subsection{Model background and purpose}

In a wide variety of ecological applications, the system that is typically under study is not homogeneously
mixing, and it is therefore necessary to address the question of how spatial interaction structure can actually affect the outcomes of evolutionary games. In this section we are therefore concerned with the influence of heterogeneous spatial structure on tag-based cooperation with four competing strategies and phenotypic diversity. 

The evolution of tag-based cooperation has thus far been studied in a wide variety of contexts but mostly with four strategies and PD-like game interactions \cite{hamaxel,laird1,hadzibeganovic1,hartshorn,hadzibeganovic2,hadzibeganovic3,Jensen2019pre,hadzibeganovic4,hadzibeganovic5}. Laird \cite{laird1} was the first to systematically investigate the dynamics of four strategies in a tag-based model with SD-game interactions. He demonstrated that in the absence of any mutation, the aspatial two-tag game tended to collapse into the traditional tagless SD game \cite{laird1}. However, in the spatial game with two tags, a 
far richer evolutionary dynamics of strategies was observed, including the cyclic coexistence of conditional cooperators and unconditional defectors. Nevertheless, the underlying interaction structure in this study \cite{laird1} was the standard regular square lattice with periodic boundary 
conditions and fully habitable lattice sites, and hence without any spatially correlated or otherwise defined structural heterogeneities such as fragmented empty sites. 

Since inhomogeneous interaction structures with non-randomly distributed empty sites have not been studied in the context of tag-based cooperation, we were interested in the present paper mainly in the dynamics of multiple strategies evolving in heterogeneously fragmented networked systems. More specifically, inspired by the work of Zhang et al. \cite{zhangetalempty}, we herein focused on the effects of spatially correlated fragmentation of empty sites on the evolution of tag-based cooperation in a structured population with phenotypic diversity and SD-game based interactions. 

\subsection{Model assumptions and parameters}

In our model, players were arranged spatially on a two-dimensional square lattice in which individual sites were characterized 
by two distinct states: empty (or unsuitable) and suitable. The empty sites in our model were indefinitely uninhabitable, i.e. they remained unsuitable for habitation throughout the system evolution. Suitable sites on the other hand represented those patches of habitat which can be occupied. Furthermore, we define the parameter $p_0$ as the proportion of empty unsuitable sites in the system, and the parameter $q_{0|0}$ as the clustering degree of empty sites, which actually stands for the first-order spatial autocorrelation of uninhabitable empty sites in the model \cite{zhangetalempty,hiebeler1}. 

Specifically, the parameter $q_{0|0}$ represents the conditional probability that a randomly selected neighbor of an unsuitable site in the lattice is also itself empty and unsuitable for habitation \cite{hiebeler1,satoiwasa,huietal2006}. Obviously, the 
proportion of \textit{habitable} sites in the model is then given by $p_{1} =1-p_{0}$, and the clustering degree of those sites that are actually suitable for habitation can be expressed as $q_{1|1}=1-\frac{p_{0}}{1-p_{0}}(1-q_{0|0})$ \cite{zhangetalempty}. Thus, the clustering degree of suitable lattice sites increases with the increasing clustering degree of unsuitable empty sites. 

Under this framework, by employing the two spatial configuration parameters, i.e. the global probability $p_{0}$ and the conditional probability $q_{0|0}$, a heterogeneous spatial structure with non-randomly distributed habitat sites, i.e. with a specific arrangement and clustering degree of unsuitable sites, can be precisely defined \cite{zhangetalempty,hiebeler1}. For instance, the heterogeneous spatial condition having $p_{0}=0.3$ and $q_{0|0}=0.3$ stands for an 
example landscape with a moderate habitat loss characterized by $30\%$ of empty sites that are distributed randomly across the lattice. The condition $p_{0}=0.3, q_{0|0}=0.1$ has the same, moderate proportion of empty sites as in the previous example, but now with a much more scattered fragmentation of sites (i.e. there are always only a few adjacent sites of the same type in the lattice). 

\begin{figure}[htb] 
\begin{center}
\hspace*{-0.1cm}
\includegraphics[scale=0.7]{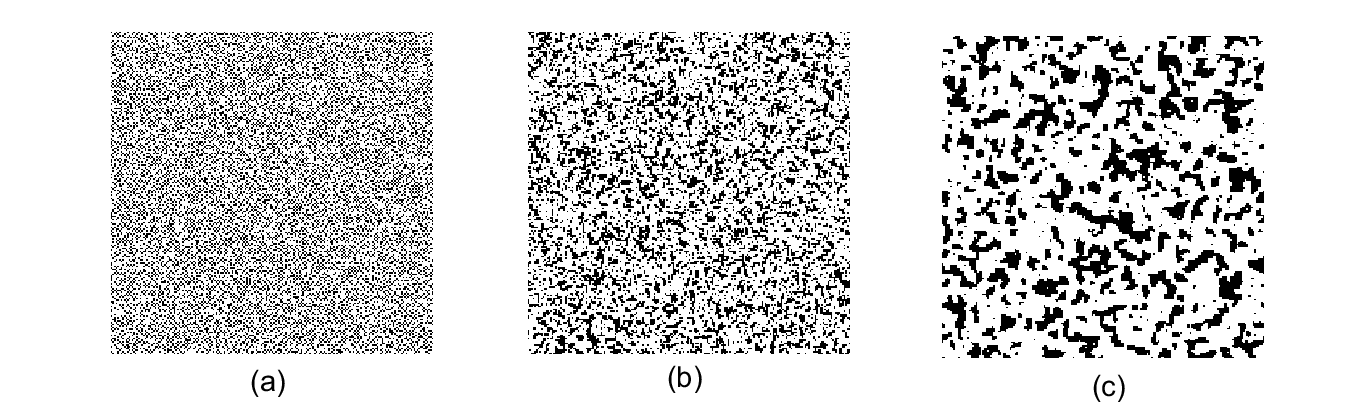} 
\caption{Examples of spatial configurations with the same degree of habitat loss ($p_{0}=0.3$) and the same system size (i.e. with $200 \times 200$ nodes), but with three different clustering degrees of empty sites, $q_{0|0} = 0.1$ (a), $q_{0|0} = 0.5$ (b), and $q_{0|0} = 0.8$ (c). Thus, all three spatial configurations have 30\% of empty sites (black), but each has a different degree of clustering, ranging from the case (a) where not so many empty sites are adjacent to each other, to the opposite case (c) where empty sites are fairly often surrounded by other empty neighbors, forming island-like empty site clusters.}
\label{fig:exampleconfigs} 
\end{center}
\end{figure}

However, although the condition $p_{0}=0.3, q_{0|0}=0.9$ has the same total amount of empty sites as the previous two examples, the probability that a randomly selected neighbor of an empty site is also empty is now much higher (and thus sites placed next to each other are in this condition more likely to be of the same type). Therefore, the unsuitable sites in the model condition $p_{0}=0.3, q_{0|0}=0.9$ are fragmented in a much more aggregate manner than in any of the previous two examples \cite{zhangetalempty,hiebeler1}, forming multiple island-like empty space barriers between suitable lattice sites (see Fig.~\ref{fig:exampleconfigs}). 

\subsection{Generation of heterogeneously fragmented empty sites}

In order to generate heterogeneously fragmented landscape structures for our computer simulations, we employed the method previously introduced by Hiebeler \cite{hiebeler1,hiebeler2}. The method is based on the pair approximation technique and basically describes the probabilities or frequencies of all 
feasible configurations for the states of $2 \times 1$ blocks of lattice sites. Here, each lattice site is characterized by a value describing a given habitat type: type 0 (unsuitable) or type 1 (suitable). Furthermore, a complete spatial symmetry is assumed such that $p_{[ij]} = p_{[ji]}$, whereby $p_{[ij]}$ represents the probability of $2\times 1$ blocks of $i$ and $j$ pairs, for $i$ and $j$ taking the values of either 0 or 1. In total, four different $2 \times 1$ block probabilities for the pairs of sites can be distinguished \cite{hiebeler1}: $p[00], p[11], p[01],$ and $p[10]$. Clearly, these four probabilities together need to add up to unity such that $p[00] + 2p[01] + p[11] = 1$ (since $p[01] = p[10]$, due to the above mentioned spatial symmetry). 

The two free parameters can then be described via the previously mentioned global probability $p_{0}$ and the conditional probability $q_{0|0}$. For example, for the two block probabilities $p[00]$ and $p[01]$, we have that $p_{0} = p[00] + p[01]$ and $q_{0|0} = p[00] / p_{0}$ \cite{hiebeler1}. It is necessary to note here that when the unsuitable and suitable habitat types are randomly arranged across the lattice, then
$q_{0|0} =  p_{0}$, because in this particular random distribution case, the conditional probability that a given neighboring site of another empty site in the lattice is also empty itself is actually equal to $p_{0}$ \cite{hiebeler1}. 

In a more general sense, for a given pair of lattice sites, the wanted i.e. desired $2\times 1$ block probabilities $p_{[ij]} (i,j=0,1)$ can be obtained through the probability relations $p_{i}=\sum_{j}p_{[ij]}$ and $p_{[ij]}=p_j p_{i|j}$. Moreover, in a similar fashion as described above, one can translate in the other direction to the $2 \times 1$ block probabilities from the global and conditional probabilities $p_{0}$ and $q_{0|0}$, i.e. $p[00]=p_{0} q_{0|0}$, $p[01]=p_{0}(1-q_{0|0})$, and from the above equation of the four probabilities summing up to unity, we have that $p[11] = 1 - p[00] - 2p[01] = 1 + p_{0}(q_{0|0} - 2)$ \cite{hiebeler1}. 

Given specific $p_{0}$ and $q_{0|0}$ values, one can then easily generate a heterogeneously fragmented landscape with a specified proportion and a specified clustering degree of empty type-0 sites via the algorithm described by Hiebeler \cite{hiebeler1}. First, given the global and the conditional probabilites $p_{0}$ and $q_{0|0}$, the wanted $2 \times 1$ block probabilities $p_{[ij]}$ are calculated using the three last equations presented above 
for $p[00], p[01],$ and $p[11]$. Next, a random lattice with the accurate proportion $p_{0}$ of type-0 sites is generated, and the \textit{measured block probabilities} $\hat p_{[ij]}$ are then obtained by checking the direct four neighbors of each individual site. In the next step, a lattice site is randomly selected and its state is designated for a change (either from 0 to 1 or from 1 to 0), provided this modification will shift the lattice's measured block probabilites more towards the desired ones; otherwise, the randomly selected lattice site will remain in its originally observed state. 

Next, we iteratively proceed with further random site selections and eventual modifications of their values, until the difference $\Delta$ between the lattice's \textit{measured} block probabilites and its \textit{desired} block probabilites, defined as $\Delta =|p_{[00]}-\hat p_{[00]}|+2|p_{[01]}-\hat p_{[01]}|+|p_{[11]}-\hat p_{[11]}|$, is smaller than a specified tolerance value (or until a predefined iteration termination threshold has been reached) \cite{hiebeler1}. This algorithm will ultimately create a heterogeneous spatial structure with empty sites containing the wanted $2\times 1$ block probabilites; all further details of this algorithm can be found in Ref. \cite{hiebeler1}. 

\begin{figure}[thb] 
\hspace*{0.05cm}
\includegraphics[scale=0.91]{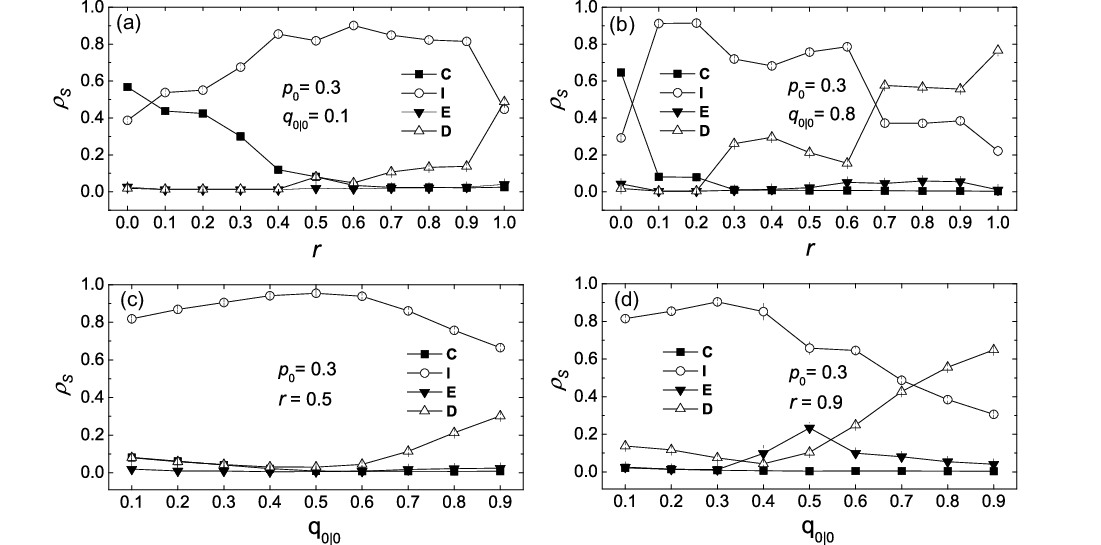}
\caption{The average equilibrium density of altruistic ($C$), ethnocentric ($I$), cosmopolitan ($E$), and egoistic ($D$) strategies in the tag-based cooperation model with heterogeneously fragmented empty sites in systems with two tags as a function of the cost-to-benefit ratio $r$ (top row panels (a)-(b)) and the clustering degree of empty sites $q_{0|0}$ (bottom row panels (c)-(d)). The results represent the averages taken over 10 independent simulation runs, each calculated over the last 5000 time steps of an at least $t=20000$ iterations long simulation run. The error bars show the corresponding standard error of the mean (SEM) values, and the lattice size for each simulation run was $100 \times 100$. The four strategies were initially equally frequent and were randomly distributed across the habitable model space. }
\label{fig:4plotstrategies}
\end{figure}

\subsection{Simulation specifications and the evolutionary process}

In the course of all simulations of our model with empty sites we implemented a synchronous updating process. Initially, players with one of the four randomly assigned strategies were distributed over habitable lattice sites with equal probability. At each time step, every individual agent occupying a habitable lattice site interacted with its four direct neighbors, and the total payoff of an agent was then calculated by summarizing the scores derived from all interactions of the same time step. We further assumed here that an individual agent obtains a zero payoff from the interaction with an empty, uninhabitable site in its neighborhood. Precise details of tag-mediated agent interactions and the corresponding payoff matrix (for two tags and four competing strategies) are given in Section 2.2. Following the interaction and payoff calculation processes, the next stage of the evolutionary process was imitation. Here, in accordance with the above described imitation mechanism from Ref. \cite{traulsen2006}, an individual agent adopts the strategy of an opponent player with the highest total payoff in its direct neighborhood. 

Unless otherwise specified, our main simulations were carried out on a square lattice with $L = 100\times 100$ sites; in addition, we tested our model on larger $L = 200\times 200$ and $L = 500\times 500$ lattices, as well as on smaller systems with $L = 50\times 50$ sites. Our main simulations were always run for a total of at least 20000 generations (or more for larger systems), and the averages were then taken over the last 5000 time steps. After obtaining the equilibrium frequencies of all competing strategies, we then analyzed the effects of varying clustering degrees of empty sites (as well as the effects of other independent variables) on the evolutionary dynamics of strategies. 

\section{Results}

Three examples of representative spatial configurations that were used in the course of our simulation experiments (all with heterogeneously fragmented landscapes containing empty sites generated with the algorithm of Hiebeler \cite{hiebeler1}), are shown in Fig.~\ref{fig:exampleconfigs}. Here, all three model configurations are characterized by the identical degree of habitat loss $p_{0}$ but distinct values of $q_{0|0}$, thus allowing us to control for the overall amount of habitat in our model while simultaneously varying the clustering degree of empty sites. To enable comparability with a related previous study \cite{zhangetalempty}, the amount of habitat loss throughout our model simulations was fixed at $p_{0}=0.3$, which was well above the previously widely assumed fragmentation-related extinction threshold \cite{fahrig2,parkermacnally}.

In Fig.~\ref{fig:4plotstrategies}, we show the average frequencies $\rho_S$ of the four strategies ($C$, $I$, $E$, and $D$) as a function of the cost-to-benefit ratio $r$ (panels (a) and (b)) for two different configurations of spatially correlated empty sites, and in panels (c) and (d) we show $\rho_S$ vs. $q_{0|0}$ for two different, fixed values of $r$. All results shown in Fig.~\ref{fig:4plotstrategies} were obtained from the model with a low phenotypic diversity, i.e. with only two tags available in the population.

At $q_{0|0}=0.1$ in a two-tag model, the maintenance of unconditional cooperation (i.e. altruism, $C$) is warranted only under lower to intermediate cost-to-benefit ratios, i.e. when $r<0.5$ (see Fig.~\ref{fig:4plotstrategies}(a)). On the other hand, in the same model condition, high levels of ethnocentric intra-tag cooperation ($I$) are maintained irrespective of the underlying value of $r$.

Moreover, at both low $r$ and low $q_{0|0}$, ingroup-biased ($I$) and unconditional ($C$) cooperation can coexist in evolutionary time in the two-tag model, with $C$ even dominating over all other strategies when $r=0$ (Fig.~\ref{fig:4plotstrategies}(a)). With an increasing $r$ but at a constantly low $q_{0|0}=0.1$, ingroup-biased ehtnocentric ($I$) cooperation is elevated, taking over the population for the vast majority of investigated cost-to-benefit ratios, i.e. for $0.0 < r < 1.0$. At $r=1$, unconditional defection, i.e. egoism ($D$) becomes substantially elevated, outweighing all other strategies. 

However, as the clustering degree of empty sites increases considerably to $q_{0|0}=0.8$ in the same system with a moderate habitat loss ($p_{0}=0.3$) and two-tags (Fig.~\ref{fig:4plotstrategies}(b)), altruism ($C$) is strongly suppressed already at $r>0$ while the elevated ethnocentric cooperation now coexists with unconditional defection $D$ at lower to intermediate levels of $r$. However, at higher cost-to-benefit ratios, defectors ultimately take over the population, i.e. as soon as $r > 0.6$. We can further observe in panels (c) and (d) of Fig.~\ref{fig:4plotstrategies} how this coexistence of $I$ and $D$ strategists, and the ultimate dominance of ethnocentrics ($I$) or egoists ($D$) in the two-tag model, depends on $q_{0|0}$ and $r$ across all considered values of the clustering degree of empty sites. 

Specifically, the unhindered dominance of ethnocentric $I$-cooperators regardless of $q_{0|0}$ is enabled in the two-tag model only at intermediate $r$, i.e. when $r=0.5$ (see Fig.~\ref{fig:4plotstrategies}(c)); however, at $r=0.5$ and $q_{0|0} > 0.6$, $I$ and $D$ strategists can coexist in the population. At higher cost-to-benefit ratios such as $r=0.9$ (Fig.~\ref{fig:4plotstrategies}(d)), this strategy coexistence is still found at greater values of $q_{0|0}$, but only under the dominance of unconditional defectors $D$ when $q_{0|0} > 0.7$ (see Fig.~\ref{fig:4plotstrategies}(d)). Together, these findings from Fig.~\ref{fig:4plotstrategies} indicate that the maintenance of altruists ($C$) in our two-tag model is not determined by either $q_{0|0}$ or $r$ alone, but it is ultimately decided via the non-trivial interaction between these two independent variables. 

\begin{figure}[thb] 
\begin{center}
\hspace*{0.9cm}
\includegraphics[scale=0.85]{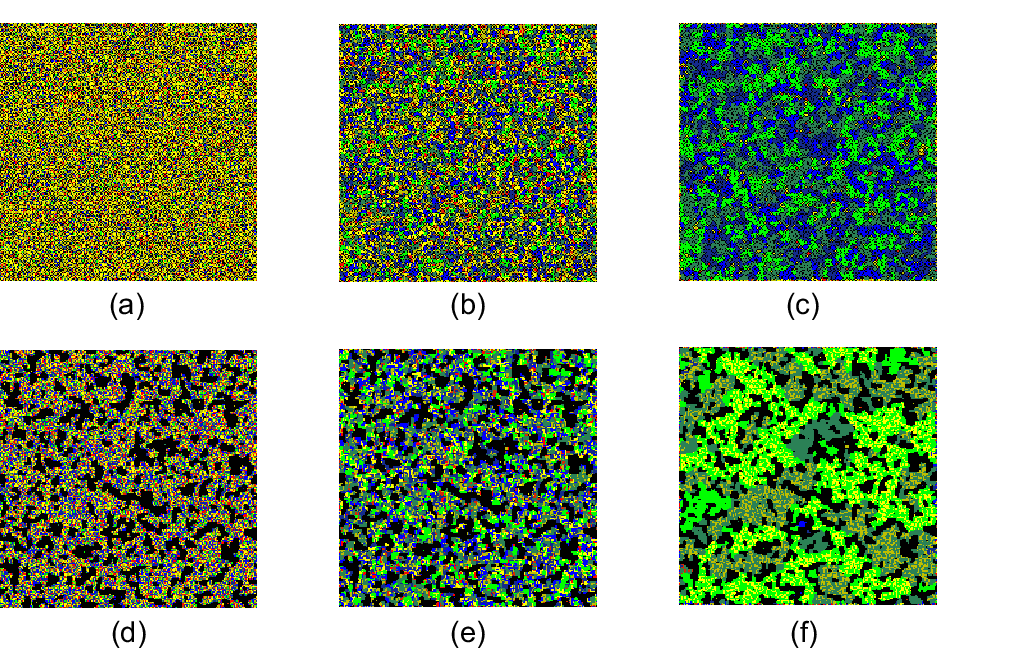} 
\vspace{-0.1pt}
\caption{Typical 2D color snapshots of the spatio-temporal dynamics of competing strategies in two-tag systems with heterogeneously fragmented empty sites. The parameter values for the shown spatial configurations were $q_{0|0}=0.1$, $r=0.3$ ((a)-(c)), and $q_{0|0}=0.8$, $r=0.3$ ((d)-(f)), and the snapshots were 
taken (from left to right) at $t=0$ ((a) and (d)), $t=10$ ((b) and (e)), and $t=20000$ time steps ((c) and (f)). The proportion of empty sites in the model was set to $p_{0}=0.3$, and the system size was $L=200 \times 200$ nodes. In this model version with four strategies and two tags (C1, I1, E1, D1, C2, I2, E2, D2), we used blue, green, red and yellow colors to represent the four strategies with the first tag (C1, I1, E1, D1), and we used dark blue, dark green, dark red and dark yellow/orange to depict the same four strategies but with the second tag (C2, I2, E2, D2). Black color always represents empty sites in the model.}  
\label{fig:2Dsnapshots1}
\end{center}
\end{figure}

Representative 2D color snapshots of the evolution of spatial distributions of strategies in a two-tag model are shown in Fig.~\ref{fig:2Dsnapshots1}. Here, the snapshots illustrate the cases of stable coexistence of $I$ and $C$ strategies (Fig.~\ref{fig:2Dsnapshots1}(c)), and $I$ and $D$ strategies in the two-tag model (Fig.~\ref{fig:2Dsnapshots1}(f)). Interestingly, while in the first condition ($q_{0|0}= 0.1$, $r=0.3$) both ethnocentrics (green) and altruists (blue) at equilibrium can build stable clusters of relatively large sizes (Fig.~\ref{fig:2Dsnapshots1}(c)), we can see that in the second 
condition ($q_{0|0}= 0.8$, $r=0.3$) only ethnocentrics (green) are able to build such stable large-sized formations (Fig.~\ref{fig:2Dsnapshots1}(f)). Egoists (yellow/orange), on the other hand, can build here only smaller but often interconnected filament-like structures, scattered within ethnocentric clusters. Their persistence in evolutionary time is largely due to their continued exploitation of ethnocentric tag-mates with whom they share the same phenotypic markers.

\begin{figure}[thb]  
\begin{center}
\hspace*{-0.6cm}
\includegraphics[scale=0.75]{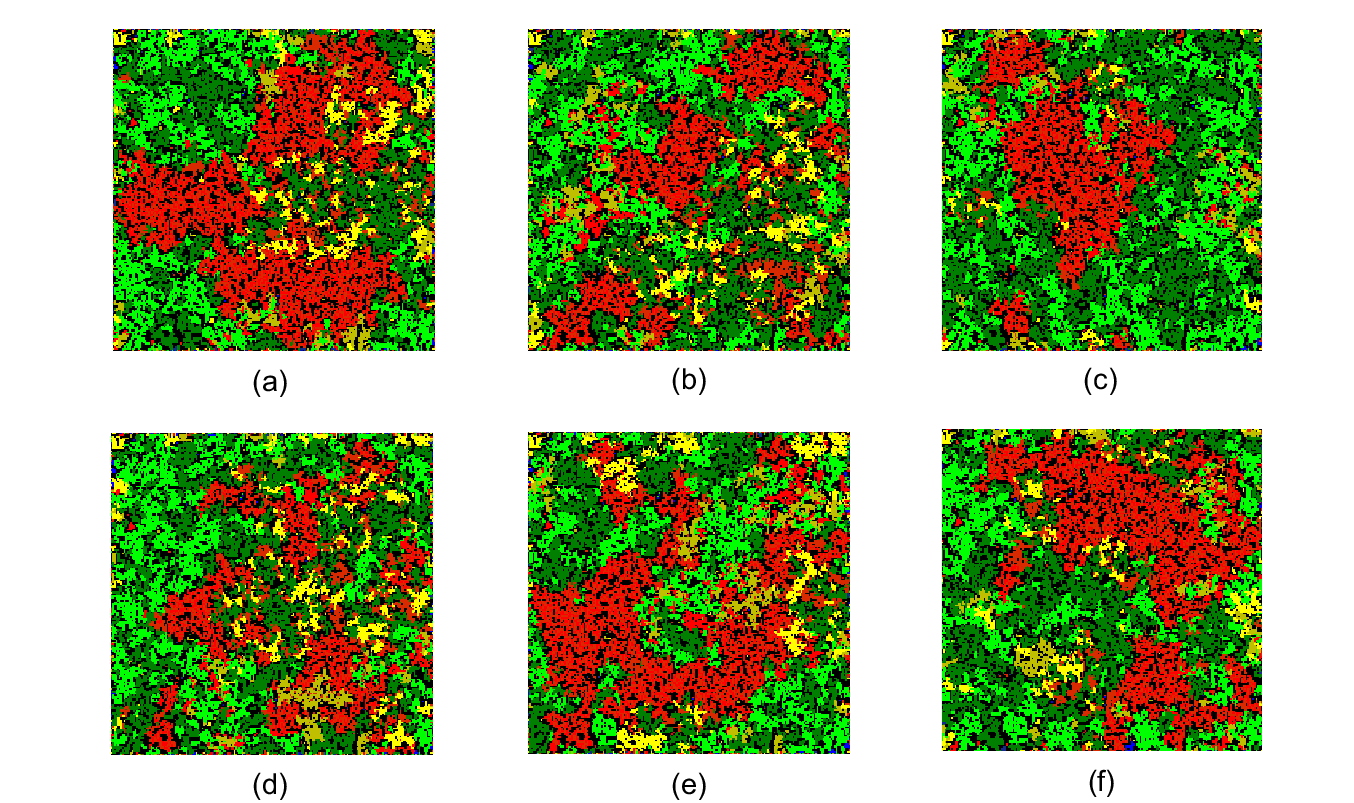} 
\vspace{-0.1pt}
\caption{Representative 2D color snapshots of the spatio-temporal dynamics of the four strategies (C, I, E, and D) in a two-tag system ($\lambda=2$) with I-E strategy coexistence. The snapshots were taken at $t=4500$ (a), $t=10000$ (b), $t=20000$ (c), $t=50000$ (d), $t=80000$ (e), and $t=100000$ time steps (f). The model parameter values were set to $p_{0}=0.3$, $q_{0|0}=0.5$, $r=0.9$, and $L=200 \times 200$. In the initial spatial configuration at $t=0$, the four strategies were randomly distributed but with initially equal strategy ratios. The coexistence of cosmopolitan extra-group cooperation (red) with ethnocentrism (green) persists in this condition even after $t=100000$ time steps (f). In this model version with four strategies and two tags, we used blue, green, red and yellow colors to represent the four strategies with the first tag (C1, I1, E1, and D1), and we used dark blue, dark green, dark red and dark yellow to depict the same strategies but with the second tag (C2, I2, E2, and D2). Black color represents uninhabitable empty sites.}
\label{fig:2Dsnapshots2}
\end{center}
\end{figure}

In Fig.~\ref{fig:2Dsnapshots2} we illustrate more closely the evolutionary dynamics of a special phenomenon which was identified in our two-tag model simulations summarized in Fig.~\ref{fig:4plotstrategies}(d). Here, we see that cosmopolitan extra-tag cooperators ($E$) can reach considerably high levels, i.e. about 1/4 of the population, thereby stably coexisting with ethnocentrics ($I$) at intermediate $q_{0|0}$ and large $r$ (see Fig.~\ref{fig:4plotstrategies}(d)). This is clearly different from our previously reported finding obtained with the aspatial model, in which $E$ strategy was always lower than the neutral case, i.e. it remained below 1/4 of the population. As evidenced in Fig.~\ref{fig:2Dsnapshots2}, this coexistence of extra-group cooperators (red) with ethnocentrics (green) appeared already in the relatively early stages of system evolution (Fig.~\ref{fig:2Dsnapshots2}(a)), persisting in this particular condition even after $t = 100000$ time steps (Fig.~\ref{fig:2Dsnapshots2}(f)). This indicates that the evolution of out-group cooperation in our two-tag model with empty sites is not simply a transient phenomenon, but it instead represents a viable alternative to ethnocentric behavior in phenotypically diverse populations. 

\begin{figure}[h] 
\begin{center}
\hspace*{-0.68cm}
\includegraphics[scale=0.65]{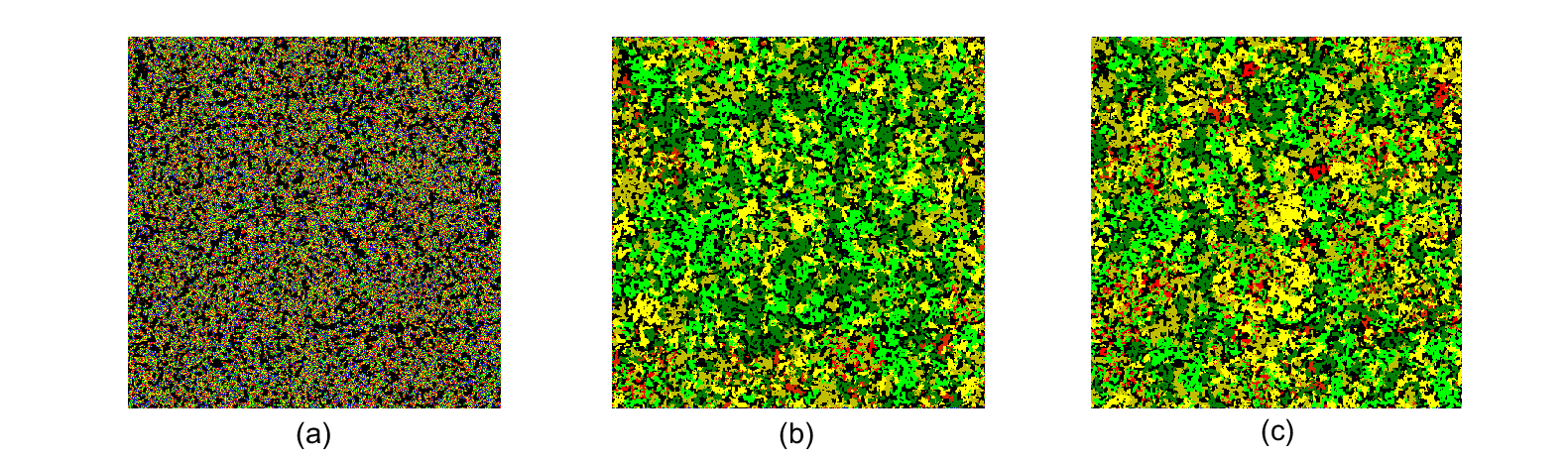} 
\vspace{-0.1pt}
\caption{2D color snapshots of the spatio-temporal dynamics of the four strategies (C, I, E, and D) in a large two-tag system with $L=500 \times 500$ (i.e. 250,000 nodes). The parameters of the model's spatial configuration were $p_{0}=0.3$ and $q_{0|0}=0.8$ and the cost-to-benefit ratio was fixed at $r=0.9$. The three snapshots were taken at $t=0$ (a), $t=1000$ (b), and $t=60000$ time steps (c). In the initial spatial configuration at $t=0$ (a), the four strategies were randomly distributed and with initially equal strategy ratios. We used blue, green, red and yellow 
colors to represent the four strategies with the first tag (C1, I1, E1, D1), and we used dark blue, dark green, dark red and dark yellow to depict the same strategies but with the second tag (C2, I2, E2, D2). Black color represents empty sites. Panel (c) shows that although egoism (D) ultimately dominates in this particular condition, where both $q_{0|0}$ and $r$ are high, the E strategy clusters (red) do not seize to exist also in much larger systems and even after $t=60000$ time steps.}
\label{fig:2Dsnapshots3}
\end{center}
\end{figure}

We further observed that these cosmopolitan $E$ clusters can survive in much larger systems and even under greater clustering degrees of empty 
sites, as shown in Fig.~\ref{fig:2Dsnapshots3}. Thus, in addition to systems with $L = 100\times 100$ sites, we tested our model on larger $L = 200\times 200$ and $L = 500\times 500$ lattices, as well as on smaller systems with $L = 50\times 50$ sites (not shown), all revealing outcomes that were qualitatively not different from those presented in Figs.~2-5.

\begin{figure} [thb]
\begin{center}
\includegraphics[scale=0.79]{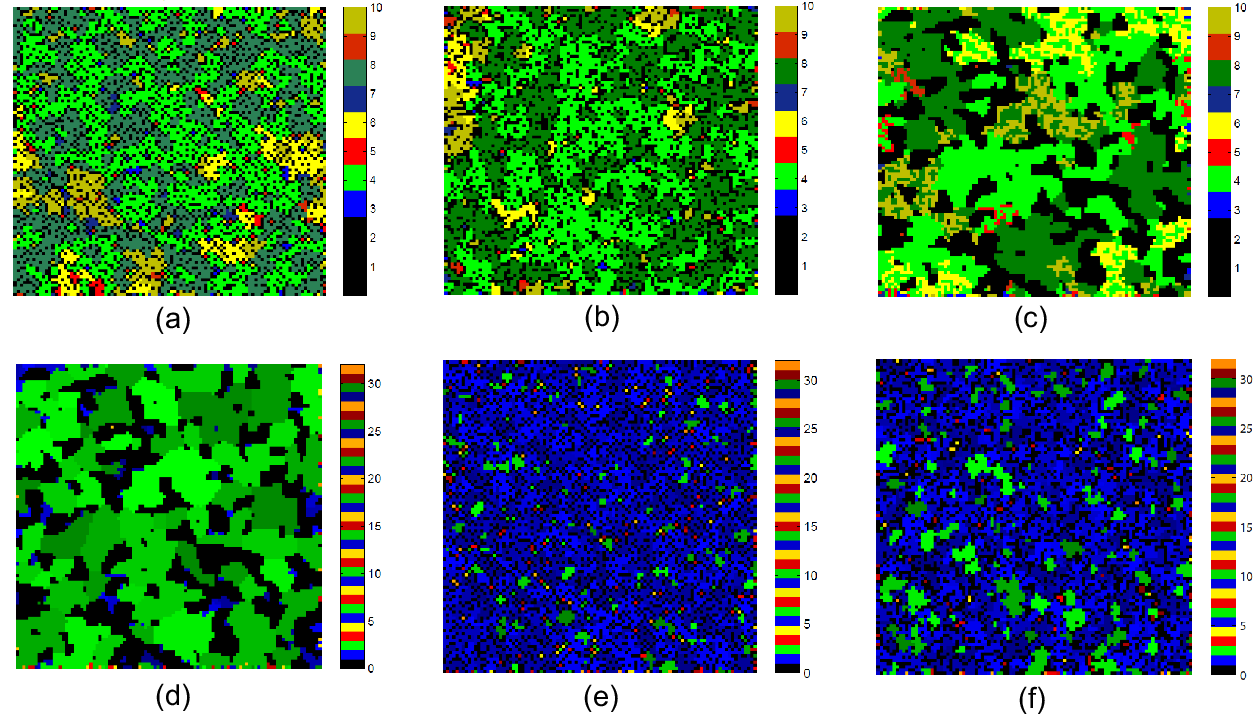}  
\vspace{-0.3pt}
\caption{Representative 2D color snapshots of spatial distributions of strategies at equilibrium in a model of the evolution of cooperation with two ((a)-(c)) and eight ((d)-(f)) tags, under varying clustering degrees of empty sites $q_{0|0}$ and different cost-to-benefit ratios $r$. The black color in all panels represents the unsuitable empty sites in the model; the different shades of green represent the $I$ strategy (ethnocentrism), different shades of blue stand for the $C$ strategy (altruism), shades of red represent the $E$ strategy (cosmopolitanism), and yellow/orange stands for $D$ (egoists). In panels (a)-(c), besides the black color (empty sites), we further have eight different colors (2 tags $\times$ 4 strategies), and in panels (d)-(f) we have 32 different colors (8 tags $\times$ 4 strategies). The proportion of empty sites in 
all shown model conditions was $p_{0}=0.3$, the lattice size was $L=100 \times 100$ nodes, and all snapshots were taken at $t=20000$. Prior to each simulation run, the four strategies were of equal frequencies, and were always randomly distributed in the initial configuration across the habitable lattice sites. The precise model conditions i.e. the combinations of $q_{0|0}$ and $r$ parameter values were: $q_{0|0}=0.1$, $r=0.9$ (a), $q_{0|0}=0.3$, $r=0.9$ (b), $q_{0|0}=0.8$, $r=0.5$ (c), $q_{0|0}=0.8$, $r=0.1$ (d), $q_{0|0}=0.1$, $r=0.1$ (e), and $q_{0|0}=0.3$, $r=0.1$ (f).}
\label{fig:2Dsnapshots4}
\end{center}
\end{figure}

To examine the effects of phenotypic diversity on the evolution of cooperation in degraded habitats with empty sites, in Fig.~\ref{fig:2Dsnapshots4} we further compared spatial distributions of strategies at equilbrium in models with two ((a)-(c)) and eight ((d)-(f)) tags, under varying clustering degrees of empty sites $q_{0|0}$ and different cost-to-benefit ratios $r$. Here, we contrasted a particularly pronounced superiority of altruism (blue) in 8-tag systems (e.g. Fig.~\ref{fig:2Dsnapshots4}(e)) against the scenario with ethnocentric (green) dominance in 2-tag systems (e.g. Fig.~\ref{fig:2Dsnapshots4}(b)).

This striking superiority of altruism in phenotypically highly diverse populations and the role that phenotypic diversity may play both individually and jointly in combination with $q_{0|0}$ and $r$ in the evolution of cooperation was investigated in more detail in Figs.~7-8. First, we can see in Fig.~7(a) that at low $r$, higher values of $q_{0|0}$ favor ethnocentrism ($I$-strategy) and lead to a reduction of altruism $\rho_C$ ($C$-strategy). We can further observe in Fig.~7(a) that for $r=0.1$, the dependence of $\rho_C$ on $\lambda$ is non-monotonic, and that superiority of $\rho_C$ over all other competing strategies is given only at $\lambda = 8$ and $q_{0|0} \leq 0.3$, but not at $\lambda < 8$ (irrespective of $q_{0|0}$). 

However, as the cost-to-benefit ratio increases from $r=0.1$ to $r=0.5$ (Fig.~7(b)), ethnocentric cooperation strongly dominates the population in all considered model conditions, i.e. regardless of phenotypic diversity or the underlying value of $q_{0|0}$. Thus, this intermediate cost-to-benefit ratio $r=0.5$ seems to be the most beneficial condition for the remarkably successful evolution of ethnocentrism in our model (Fig.~7(b)). However, as the cost-to-benefit ratio is further elevated to $r=0.9$ (Fig.~7(c)), ethnocentric cooperation remains dominant in the models with four and eight tags over the whole range of values of $q_{0|0}$, but in the phenotypically less diverse model (with only two tags), ethnocentrism decreases with an increasing $q_{0|0}$ and is ultimately taken over by defectors $D$ at a higher clustering degree of emtpy sites such as $q_{0|0}=0.8$. Thus we can see from the three panels of Fig.~7 that the moderating effect of $\lambda$ on the relationship between $q_{0|0}$ and $\rho_S$ changes as a function of $r$. We therefore examined a potential three-way interplay between $\lambda \times q_{0|0} \times r$ in relation to the resulting mean frequency of strategies $\rho_S$ in our model (Fig.~7). Thus, the independent variables in this $3 \times 4 \times 3$ factorial design were the number of tags $\lambda$ (2, 4, and 8), the clustering degree of empty sites $q_{0|0}$ (0.1, 0.3, 0.6, and 0.8), and the cost-to-benefit ratio $r$ (0.1, 0.5, and 0.9). 

We can observe the form of this complex interaction in Fig.~7. First, we can confirm that the two-way interaction between $\lambda \times q_{0|0}$ on the evolution of altruists $\rho_C$ indeed varies with the level of $r$. Specifically, at low $r$ ($r=0.1$), the slope of the curve representing the association between $q_{0|0}$ and $\rho_C$ (frequency of altruism) was steeper under the high phenotypic diversity ($\lambda=8$) than under low ($\lambda=2$) or intermediate ($\lambda=4$) number of tags (see Fig.~7(a)). This difference in slopes suggests that when $r$ was low but $\lambda$ high, the level of $q_{0|0}$ was strongly negatively related to the frequency of altruism $\rho_C$. However, when the clustering degree of empty sites increased substantially to $q_{0|0}=0.8$, $\rho_C$ did not depend on $\lambda$ or $r$ . 

Importantly, this difference in slopes and the associated negative relationship between $q_{0|0}$ and $\rho_C$ observed at $r=0.1$ was retained to a rather small extent when $\lambda=8$ at $r=0.5$ (Fig.~7(b)), but was completely lost when the cost-to-benefit ratio further increased to $r=0.9$ (Fig.~7(c)), where the level of altruism was now at its lowest, irrespective of $\lambda$. Overall, the highest level of altruism in our model (close to 90\% of the population) was found when both $q_{0|0}$ and $r$ were low but simultaneously $\lambda$ was high (see Fig.~7(a)). 

Another similarly non-trivial interplay between $\lambda \times q_{0|0} \times r$ was observed in relation to the mean frequency of ethnocentrism $\rho_I$. At low $r$ ($r=0.1$), the slopes of the lines representing the association between $q_{0|0}$ and $\rho_I$ were now positive (see Fig.~7(a)). The level of $q_{0|0}$ was now strongly positively related to the frequency of ethnocentrism $\rho_I$ when $r$ was low but $\lambda$ high; however, $\rho_I$ did not significantly depend on $\lambda$ at $r=0.1$ when the clustering degree of empty sites increased to $q_{0|0}=0.8$. Moreover, this strong positive relationship between $q_{0|0}$ and $\rho_I$ at $r=0.1$ (Fig.~7(a)) diminished with an increasing $r$ and changed even to a negative relationship between the clustering degree of empty sites and the frequency of ethnocentrism when $r=0.9$ and $\lambda=2$, but not at an intermediate or high number of tags in the model (Fig.~7(c)). Overall, the highest level of ethnocentrism in our model was seen when the level of $q_{0|0}$ was rather intermediate ($q_{0|0}=0.6$), and simultaneously, both $r$ and $\lambda$ were high. The lowest level of ethnocentrism, on the other hand, was observed when both $q_{0|0}$ and $r$ were low but $\lambda$ was high, which also corresponded to the condition required for reaching the highest level of altruism in our model. 

Finally, the highest level of egoism in our model was observed when $q_{0|0}$ and $r$ were both high but $\lambda$ was low, and the lowest egoism was seen when $q_{0|0}$ and $\lambda$ were high while simultaneously $r$ was low. 

The aforementioned three-way interplay $\lambda \times q_{0|0} \times r$ can also be confirmed when looking at Fig.~8, which now shows the frequencies of strategies $\rho_S$ as a function of $r$ for the four different levels of $q_{0|0}$ and three different numbers of tags $\lambda$ in the model. Here, the negative relationship between the cost-to-benefit ratio $r$ and the frequency of altruism $\rho_C$ increases with the lower $q_{0|0}$ and the higher $\lambda$. We can observe substantially greater slope differences among the curves representing the relationship between $r$ and $\rho_C$ when $\lambda=8$ relative to the cases 
where $\lambda \leq 4$. Also, the association between $r$ and $\rho_I$ and the corresponding slope differences among the curves at different $q_{0|0}$ change significantly with an increasing $\lambda$, establishing now a more positive relationship between the frequency of ethnocentrism and the cost-to-benefit ratio. 

Interestingly, we can observe that at intermediate and high $\lambda$, the curves associating $r$ with $\rho_C$ and $r$ with $\rho_I$ mirror each other at almost all different levels of $q_{0|0}$ (see Fig.~8(b)-(c)). However, this overall 'mirror effect' is lost at $\lambda=2$, i.e. it is roughly preserved only for the two lowest levels of $q_{0|0}$.


\begin{figure}[htb] 
\begin{center}
\hspace*{-0.1cm}
\includegraphics[scale=0.78]{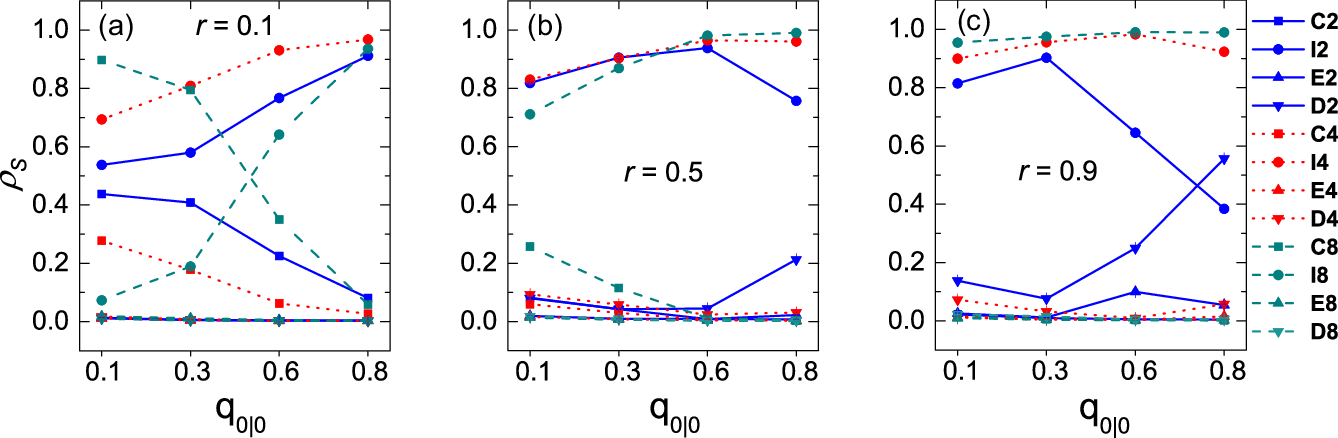}
\vspace{-0.3pt}
\caption{The stationary fractions of strategies as a function of the clustering degree of empty sites $q_{0|0}$ for the systems with different numbers of tags (2, 4, and 8), and three distinct cost-to-benefit ratios $r=0.1$ (a), $r=0.5$ (b), and $r=0.9$ (c). The proportion of empty sites was $p_{0}=0.3$, and the lattice size was always $L= 100 \times 100$ nodes. The shown data represent the averages taken over 10 independent simulation runs; the error bars represent the standard error of the mean (SEM). For each independent run, simulations were conducted for at least $t=20000$ generations, and then averages were taken over the last $t=5000$ time steps. The starting configuration of each simulation (at $t=0$) consisted of initially equal frequencies of the four strategies (C, I, E, and D) that were randomly distributed across the habitable lattice sites. Blue, red and dark cyan colors correspond to the systems with 2, 4 and 8 tags, respectively. }
\end{center}
\end{figure}

\begin{figure}[htb] 
\begin{center}
\hspace*{-0.1cm}
\includegraphics[scale=0.78]{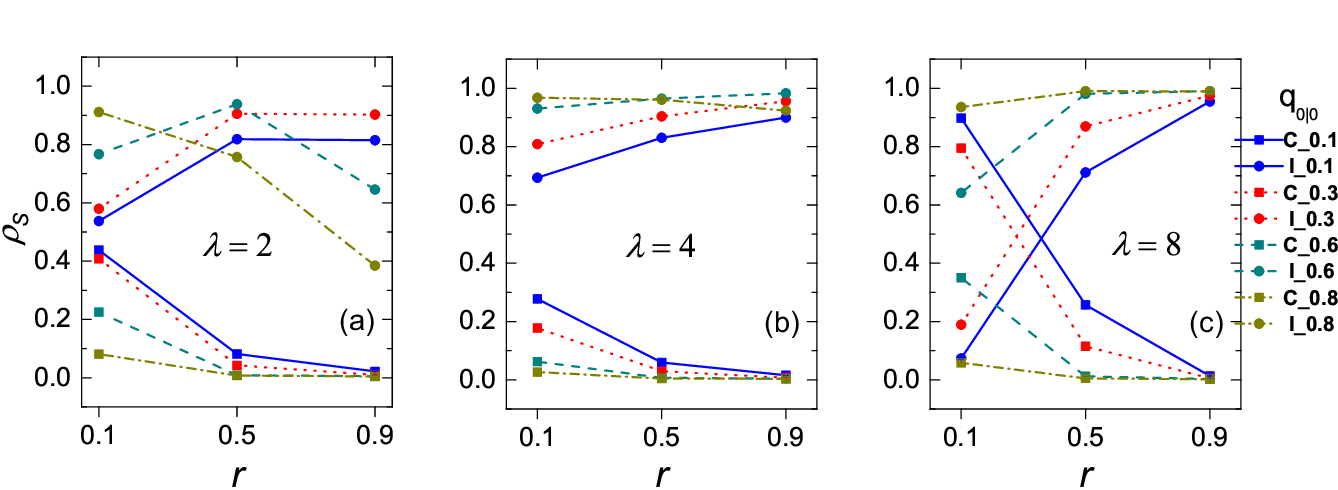}
\vspace{-0.3pt}
\caption{The stationary fractions of C and I strategies as a function of the cost-to-benefit ratio $r$ for the systems with distinct clustering degrees of empty sites $q_{0|0}$, and three different numbers of tags in the population, $\lambda=2$ (a), $\lambda=4$ (b), and $\lambda=8$ (c). The proportion of empty sites was $p_{0}=0.3$, and the lattice size was always $L= 100 \times 100$ nodes. The data represent the averages taken over 10 independent simulation runs; the error bars represent the standard error of the mean (SEM). Simulations for each independent run were conducted for at least $t=20000$ generations, and then averages were taken over the last $t=5000$ time steps. The starting configuration of each simulation (at $t=0$) consisted of initially equal frequencies of the four strategies that were randomly distributed across the habitable lattice sites. Solid squares represent altruism (C) and solid circles stand for ethnocentrism (I). Blue, red, dark cyan, and dark yellow colors correspond to the conditions with four different clustering degrees of empty sites $q_{0|0} = 0.1, 0.3, 0.6, 0.8$. }
\end{center}
\end{figure}

\begin{figure*}[hbt]
  \begin{center}
  \hspace*{-1cm}
\includegraphics[scale=0.89]{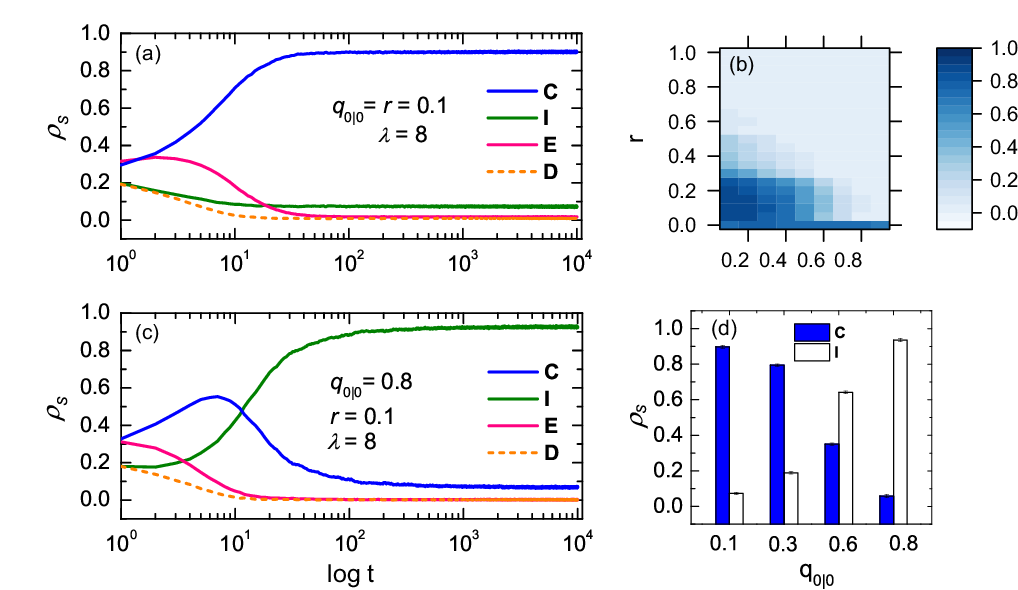} 
    \vspace{0.2cm}
    \caption{Distinct effects of the two model parameters, $q_{0|0}$ and $r$, on the evolution of strategies in systems with high phenotypic diversity ($\lambda=8$ tags). The panel (a) shows the evolutionary dominance of altruism (C) over all other competing strategies in a system with $\lambda=8$ tags when $q=r=0.1$. The panel (b) shows the heatmap of the normalized difference $\Delta_{\mathrm{C,I}}$ between altruists and ethnocentrics in the plane of model parameters $r - q_{0|0}$ for a system with $\lambda=8$ tags; the bottom-left area of the $r - q_{0|0}$ plane, with dominant dark-blue color (at low $r$ and low $q_{0|0}$ values), corresponds to the region of the parameter space for which the altruistic strategy always wins over all other competing strategies in the model. The panel (c) depicts a condition with a high clustering degree of empty sites and low $r$, in which the ethnocentric strategy I takes over the population. The panel (d) shows the average frequencies of C and I strategies at different clustering degrees of empty sites $q_{0|0}$ when $r=0.1$ and $\lambda=8$; error bars correspond to SEMs. The degree of habitat loss for all shown results was always $p_{0}=0.3$ and the lattice size was $L = 100 \times 100$. Prior to each simulation run, the four strategies C, I, E, and D were randomly distributed across the habitable lattice sites but always with initially equal strategy ratios, i.e. as 1:1:1:1.}
    \label{test4}
  \end{center}
\end{figure*}

\begin{figure}[htb]
\begin{center}
\hspace*{-0.75cm}
\includegraphics[width=15cm]{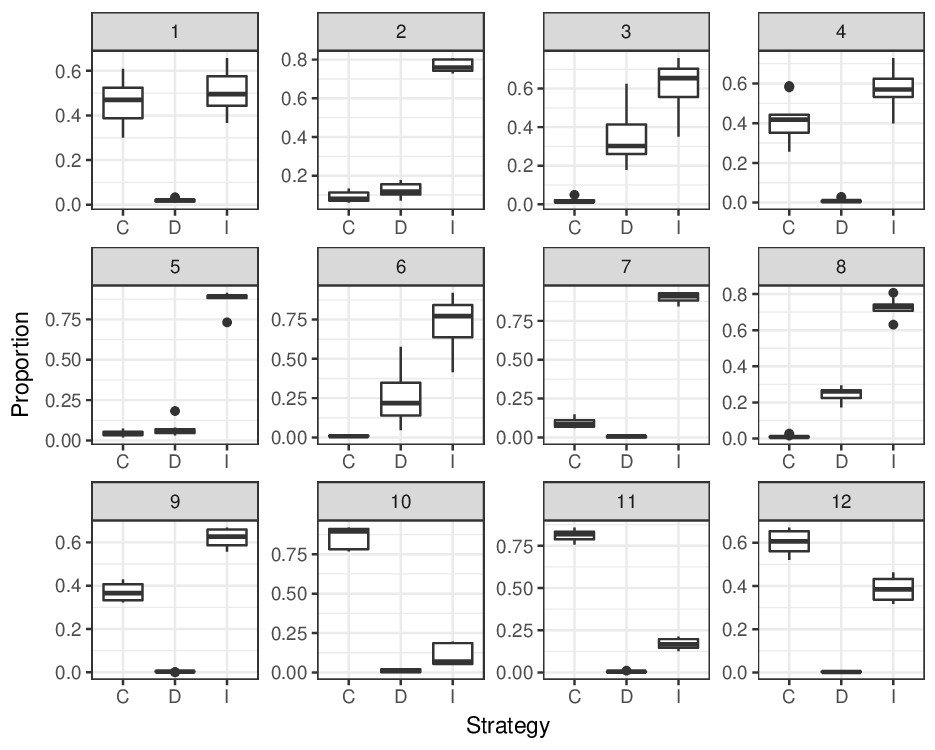}\quad
\caption{Variability of individual strategies (C, D, and I) resulting from initially different strategy ratios across 12 unique model conditions, each with a distinct combination of the model parameters $q_{0|0}$ and $r$ and different numbers of tags. The panels 1-8 in the top two rows show the results for 2-tags systems with the following parameter values: $q_{0|0}= 0.1, r=0.1$ (1), $q_{0|0}= 0.1, r=0.5$ (2), $q_{0|0}= 0.1, r=0.9$ (3), $q_{0|0}= 0.3, r=0.1$ (4), $q_{0|0}= 0.3, r=0.5$ (5), $q_{0|0}= 0.3, r=0.9$ (6), $q_{0|0}= 0.8, r=0.1$ (7), and $q_{0|0}= 0.8, r=0.5$ (8). The bottom row of panels (9-12) shows the results for the 8-tags systems, with the parameter values: $q_{0|0}= 0.6, r=0.1$ (9), $q_{0|0}= 0.1, r=0.1$ (10), $q_{0|0}= 0.3, r=0.1$ (11), and $q_{0|0}= 0.5, r=0.1$ (12). The strategy E is not shown as it is usually vanishingly low under most shown conditions. Each box in each of the 12 panels captures the variation of a given strategy across 10 independent simulation runs, each conducted with initially different ratios of the four strategies (i.e. instead of the uniform initial distributions of strategies 1:1:1:1, as in our main simulations, other initial strategy ratios were employed such as 1:1:1:2, 1:2:1:1, 1:1:2:2, 1:2:1:2, 1:1:1:5 etc., with numbers representing the ratios of the four strategies C:I:E:D at $t=0$). }
\end{center}
\end{figure}

In Fig.~9 we show the typical time series for the fractions of four competing strategies ($C$, $I$, $E$, and $D$) in the 8-tag model for the conditions resulting in altruistic (Fig.~9(a)) versus ethnocentric (Fig.~9(c)) dominance. In Fig.~9(b) we present the normalized difference $\Delta_{\mathrm{C,I}}$ between altruists $C$ and ethnocentric cooperators $I$ in the plane of model parameters $r - q_{0|0}$ for a system with $\lambda=8$ tags, where this normalized difference was defined as $\Delta_{\mathrm{C,I}} = (C-I) / (C+I)$. Fig.~9(b) captures the region of the parameter space (in the bottom-left corner) for which altruists outcompete ethnocentric cooperators and are effectively ahead of all other competing strategists (this region is highlighted in dark blue colors). Fig.~9(d) depicts the average frequencies of $C$ and $I$ strategies at different $q_{0|0}$ when $r = 0.1$ and $\lambda = 8$. It summarizes how ethnocentric agents take over the population with an increasing $q_{0|0}$, whereas altruists dominate at lower to intermediate clustering degrees of empty sites.

We further sought to ensure that the previously observed significant differences between strategies in our model were robust to also stronger types of variation in the intial conditions, i.e. beyond the standard variability obtained using different random number seeds for the initial system configurations. To address our model's sensitivity to variable initial conditions, besides using random number seeds, we also varied the respective initial ratios of the four strategies (C:I:E:D), that were set at the initialization stage prior to the start of each inidividual simulation. For example, instead of using 1:1:1:1 ratios of strategies (where the four strategies were all distributed in equal proportions in the initial configuration, as was done in most of our previous simulations reported in the present paper), we now employed a variety of strategy ratios with different initial proportions of strategies at $t=0$, such as 1:1:1:2, 1:2:1:1, 1:1:2:2, 1:2:1:2, 1:1:1:3, 1:1:1:4, 1:1:2:4, 1:1:4:4, 1:1:7:1 and 1:1:1:5, where thus strategies other than altruism were initially 
more frequently present in the population. 

As shown in Fig.~10, this resulted in a remarkably low variability of outcomes for a wide variety of model conditions and even under radically different initial strategy configurations. For example, in panels 10-12 of Fig.~10, we see that altruism ($C$-strategy) remained stably superior even if other strategies were initially significantly more often found in the population than $C$, e.g. even when there were four or five times more defectors than altruists in the initial configuration, or when both extra-group cooperators and defectors outnumbered $C$ strategists in the beginning of the simulation. Admittedly, however, in some conditions (e.g. panels 3 and 6 in Fig.~10) the model was prone to a greater variability in strategies than in others (e.g. panels 7 and 11 in Fig.~10).

Finally, Appendix B gives typical times series of the fractions of the four competing strategies in our model. We see there how the difference between $C$ and $I$ strategies changes as a function of the different variables. Moreover, we see in Figs.~B11(e)-(f) two unique conditions for which we identified characteristic oscillations of strategies $I$, $E$, and $D$. We found that these oscillations were present only in 2-tag systems, and were rather consistently quasi-periodic. These oscillations did not cease to exist even after $10^6$ time steps, and did not appear at this magnitude in any other investigated conditions in our model. Moreover, we note that our model simulations with varying initial strategy ratios did not result in significantly different outcomes relative to those reported previously with initially equal strategy proportions, even under strong strategy oscillations. 
         
\section{Discussion and future research directions}

We developed an agent-based model of tag-based cooperation with explicitly designed heterogeneous fragmentation of degraded habitats, in which we investigated the combined effects of spatially correlated empty sites, phenotypic diversity, and varying cost-to-benefit ratios $r$ on the evolution of cooperation in structured populations with multiple competing strategies. Agents in our model engaged into pairwise Snowdrift games and were able to adopt co-players' strategies via imitation. The amount of habitat loss and the heterogeneous fragmentation of sites unsuitable for habitation were operationalized in our model via two parameters: the proportion $p_{0}$ of empty sites and their clustering degree $q_{0|0}$, i.e. the conditional probability that the neighbor of an empty site is also empty itself and unsuitable for habitation. Together, these two parameters enabled generation of a variety of heterogeneous spatial configurations that served as interaction structures in our model simulations, allowing us to systematically investigate the effects of variably fragmented habitat sites on cooperation while simultaneously controlling for the total amount of destructed habitat.

Our model was tested on both smaller and larger systems, ranging from 2500 up to a quarter million of lattice sites. We observed no significant system size effects or any substantial effects of the variation in initial conditions, unless the ratio of defectors was drastically larger relative to all other competing strategies.

In addition to the observed individual effects of the studied idependent variables on cooperation, we found that due to their complex interplay, the equilibrium density of individual strategies in our model also depended nontrivially on the specific combinations of levels of all three variables. Perhaps most remarkably, at simultaneously low to intermediate clustering degrees of empty sites $q_{0|0}$ and high phenotypic diversity $\lambda$, pure altruism was observed to outweigh all other competing strategies but only in a model with low $r$ and not at intermediate or high cost-to-benefit ratios. Thus, the moderating effect of phenotypic diversty $\lambda$ on the relationship between $q_{0|0}$ and $\rho_C$ changed significantly depending on 
the level of $r$. Such a strong prevalence of pure altruism with a simultaneously radical suppression of other competing strategies was previously reported to appear in only a few tag-based cooperation models (e.g. \cite{hadzibeganovic1,hadzibeganovic4}), none of which, however, involved spatially explicit environmental heterogeneities with clustered empty sites. 

However, in our phenotypically diverse model with eight tags at $r = 0.1$, pure altruism was hindered in favor of ethnocentrism if the clustering degree $q_{0|0}$ exceeded moderate levels, forming more aggregate, island-like groups of empty sites. This facilitation of ethnocentrism at higher $q_{0|0}$ in the 8-tag model was marked at first by the coexistence of ethnocentric $I$-strategists with unconditional $C$-cooperators (at $q_{0|0}=0.6$), but was also characterized by an almost complete elimination of pure altruism at $q_{0|0}=0.8$. More generally, cooperation (either conditional or unconditional) in our model prevailed at rather low to intermediate cost-to-benefit ratios $r$ in most of our considered model versions. However, as $r$ and $q_{0|0}$ increased to considerable levels, defection became much more pronounced, especially in the two-tag model where it was able to ultimately take over the population. 

As perhaps the most notable observed effect of a single independent variable on the evolution of individual strategies in our model, we highlight the influence of intermediate cost-to-benefit ratio $r$. Here, we found a remarkable advantage of ethnocentric cooperation which at $r=0.5$ outcompeted all other strategies in all considered model versions, regardless of the number of tags and irrespective of the underlying clustering degree of empty sites. This finding is corroborated by earlier studies reporting striking robustness of ingroup-biased strategies such as ethnocentrism in computational agent-based models with four competing strategies \cite{hamaxel,hartshorn}, but it also highlights the existence of rather fragile conditions under which ethnocentric cooperation can be endangered and potentially controlled in favor of other socially more desirable strategies such 
as altruism \cite{laird1,hadzibeganovic1,hadzibeganovic2,hadzibeganovic3,ramazi,hadzibeganovic4,jeongetal}. 

For example, in Ref. \cite{hadzibeganovic2}, ethnocentric strategy was weakened in the presence of within-system migratory behaviors, but only if the migration radius, i.e. the length of movement from the original location to a new destination was sufficiently long. Moreover, in a system with both internal migrations and immigration, ethnocentric cooperation was further reduced with the increasing number of simultaneously immigrating, external individuals who invaded the system from the outside. However, in spite of these migratory behaviors and the associated measurable reductions in ethnocentric cooperation, ethnocentrism still remained relatively stable prevailing consistently in a variety of mobile systems with four competing strategies \cite{hadzibeganovic2}. 

More pronounced reductions of ethnocentrism and facilitation of other, non-ethnocentric cooperative strategies have been observed recently in tag-based systems with reproductive skew and demographic fluctuations \cite{hadzibeganovic5} in which ethnocentrism coexisted with altruism and other strategies but was not significantly ahead of them, and in populations with time-varying tags, six competing strategies, and heterogeneous immigration dynamics \cite{jeongetal}, in which ethnocentric cooperation was reducible to vanishingly low levels. We have seen in the present paper that such significant reduction and control of ethnocentrism can also be reached in phenotypically more diverse populations at preferably lower cost-to-benefit ratios in combination with heterogeneously fragmented empty locations, i.e. sites unsuitable for habitation that are distributed in a non-aggregate, scattered manner throughout the environment. 

Aggregate distributions of empty sites (e.g. $q_{0|0}=0.8$) cannot thus promote altruism, but they can support and even further enhance the cluster formation of ethnocentric $I$-agents. However, a remarkable preservation of altruism in our model with empty sites can also be observed in the condition in which $I$-strategy dominates over all others on suitable sites placed among many aggregate, island-like empty regions, thanks to which small clusters of altruism can survive in evolutionary time and resist the invasion of ethnocentrism. Thus, at lower cost-to-benefit ratios $r$, the aggregate structure of empty sites in our model can permanently protect altruistic clusters from extinction, thereby enabling their stable coexistence with ethnocentrism.  

Interestingly, previous investigations suggested that strategy coexistence in evolutionary games is a rather transient phenomenon typically limited to only certain types of games \cite{laird1,hauertholmes}, and that unconditional and conditional strategies should not actually coexist together in tag-based cooperation models since they are normally mutually exclusive \cite{masudajtb}. However, while using the Snowdrift game as a paradigmatic example for agent interactions in our heterogeneous network model with unsuitable sites, the 
reported cases of strategy coexistence (either between $I$ and $D$ or $I$ and $C$ strategies) were never found to be only transient but instead they were observed as rather stable phenomena that persisted over considerably long periods of evolution in a number of model conditions, including all three levels of phenotypic diversity (two, four, and eight tags) and at varying clustering degrees of empty sites. 

Thus, in contrast to the study in Ref. \cite{masudajtb}, the case of stable coexistence between conditional $I$-strategy and unconditional $C$-strategy in our model with empty sites further confirmed that ethnocentrism and altruism are not necessarily always mutually exclusive and can therefore stably coevolve within the same model of tag-based cooperation. Notably, only a few recent studies have found such stable coevolutionary dynamics of pure altruism and ethnocentrism, reporting several potent mechanisms responsible for this type of coexistence in tag-mediated cooperation models, including joint effects of interaction structure and memory \cite{hadzibeganovic1}, high mutation rate and low $r$ in structured populations \cite{laird1}, intergroup mating \cite{mourishultz}, nonconformity in spatial games \cite{hadzibeganovic4}, and heterogeneous reproductive success \cite{hadzibeganovic5}. Our current investigation thefore contributes to this special research line by identifying the importance of combining heterogeneous empty site fragmentation with phenotypic diversity as a novel mechanism for the coexistence of conditional and unconditional strategies in spatially structured systems with tag-mediated interactions. 

Another non-transient phenomenon observed in our study was the fixation of cosmopolitan extra-tag cooperators ($E$) at considerably
high levels and their coexistence with ethnocentric ($I$) agents at intermediate $q_{0|0}$ and large $r$ (see Fig. 2(d)). This is in contrast with our previously reported finding obtained with the aspatial model version, in which the extra-tag cooperation strategy was always lower than the neutral case, i.e. it remained below 1/4 of the population. We note that elevated extra-tag benevolence is a relatively rare phenomenon in evolutionary models of tag-based cooperation, thereby rightfully deserving its further systematic investigation and analysis. 

In addition to the spatial model, which due to its analytical intractability was only studied by means of agent-based computer simulations, we also developed analytic arguments for strategy transitions in a simplified, aspatial tag-based cooperation model with $K$ tags and four different strategies, evolving in infinite, well-mixed populations under the constraint of weak selection. Given the obtained transition rates, we further established a transition matrix between the four competing strategies in our aspatial model with $4K$ types of individuals, that holds for very large populations and weak selection pressures. Our analysis revealed conditions for the two different rankings of strategists in our aspatial tag-based cooperation model, with the relative rankings of $C$ and $D$ strategists always depending on model parameters. 

Moreover, we found that the fraction of $I$ strategy was always larger than 1/4, indicating superiority of ethnocentrism in aspatial models under many different conditions. This result confirms that tag-based cooperation can indeed emerge also in aspatial systems, in the virtual space of phenotypic features and without any explicit interaction structure, as has been shown previously in both theoretical and empirical studies \cite{antal,hadzibeganovic4,efferson}. However, even though it predicted a considerably robust dominance of ethnocentric cooperation, our aspatial model was clearly not able to capture the possible dominance of pure altruism that was discovered only thanks to extensive agent-based computer simulations of our spatially explicit model of tag-based cooperation with precisely structured environmental heterogeneities.

Importantly, previous studies \cite{alizon,sekiguchi,yangphysa,yangpla} have shown that intermediate amounts of empty sites or moderately destructed habitats can indeed promote indiscriminate cooperation. However, these earlier studies did not address the question of how the habitat structure actually needs to be distributed in a landscape in order to optimally promote global cooperation not just under two-strategy scenarios, but also in the presence of several unconditional and conditional strategies. They thus did not consider structured heterogeneities of empty sites, and were mostly conducted in the context of either PD games \cite{yangphysa} or PG games \cite{yangpla} with unconditional strategies. In addition, the model in Ref. \cite{alizon} actually assumed that empty sites can give rise to more 'free space' in the habitat which can in turn be used for the newly produced offspring. However, this was clearly different from our study with SD game interactions in which empty sites were uninhabitable throughout the simulation, representing thus unsuitable sites that are permanently unfit for habitation. 

The study of Zhang and colleagues \cite{zhangetalempty} was the first one to explicitly include these structured environmental heterogeneities with empty sites and to investigate their influence on the evolution of strategies in both PD and SD games, finding remarkable facilitatory effects of autocorrelated empty sites on cooperation in PD games, but variable effects in the SD game depending primarily on the benefit $b$ of cooperation. For example, in their model with SD game interactions and heterogeneously structured emtpy sites \cite{zhangetalempty}, cooperation evolved at higher $q_{0|0}$ but only if temptation to defect $b$ was sufficiently high, whereas at intermediate and lower values of $b$ cooperation was prevented from fixation. In our model, on the other hand,  cooperation (either conditional or unconditional) dominated at rather low to intermediate cost-to-benefit ratios $r$ in most of our considered model versions, while intra-tag cooperation could stably evolve at higher $q_{0|0}$ and many different values of $r$ (e.g. in the four tag model). 

However, it needs to be stressed here that the study of Zhang and colleagues \cite{zhangetalempty} considered only a tagless model of cooperation with two unconditional and one conditional strategy, and was therefore not only strategy-wise different from ours, but it also contained a phenotypically purely homogeneous population of individuals. In other words, prior to our present study, it was not well understood how the variation in empty site distribution under same total amounts of unsuitable sites can affect the evolution of multiple conditional and unconditional strategies in a tag-based model of cooperation, and under precisely what ecological conditions in the presence of uninhabitable sites can altruism and strategies other than ingroup-biased ethnocentrism thrive and prevail in phenotypically diverse populations. Our present investigation is therefore the first one to systematically address the effects of spatially correlated environmental heterogeneities on the evolution of multiple strategies in the context of tag-based cooperation models with fragmented habitats. 

Our findings are partly corroborated by a more recent study showing that when a population is strongly fragmented into a number of smaller groups, it becomes much more challenging to overcome group-bound cooperative behavior \cite{grossetal2023}; however, beyond confirming this finding, our results have also demonstrated that overcoming ingroup-biased cooperation and transiting towards universal benevolence can still be attained successfully in a phenotypically diverse population in the presence of unsuitable sites that are heterogeneously distributed, provided the cost-to-benefit ratio is sufficiently low, the overall degree of habitat loss is moderate, and the clustering degree of empty sites that separate the fragmented habitat remnants is not too large.

Throughout our model simulations, the amount of habitat loss remained fixed at $p_{0}=0.3$, which was well above the theoretically observed fragmentation-related extinction threshold \cite{fahrig2,parkermacnally}. Thus, it has been reported in earlier studies that the influence of habitat fragmentation depended critically on the overall amount of remnant habitat, such that only above a certain threshold of habitat loss (usually above 70\% of lost habitat), the population extinction probability increased considerably with an increasing habitat fragmentation \cite{fahrig2,parkermacnally}. 

In the present study, we observed significant effects of habitat fragmentation even if our model landscapes contained 70\% of remnant habitat (i.e. only 30\% of habitat loss), suggesting that habitat fragmentation can actually exhibit its significant effects even if preserved habitat fraction is far away from the previously assumed critical threshold. Our findings are, however, in line with a more recent study \cite{NatEcolEvol2024} showing that habitat fragmentation at a low amount of habitat loss is indeed conducive to stable coexistence phenomena. 

Clearly, future extensions of our current model should also more thoroughly examine potential interaction effects between habitat loss and habitat fragmentation \cite{parkermacnally}, which was, however, beyond the scope of the present study as we intentionally studied the effects of fragmentation while controlling for the amount of habitat loss in the model. For the combined effects of $p_{0}$ and $q_{0|0}$ on the evolution of cooperation in a related but simpler model with three strategies without tags, the reader is referred to Ref. \cite{zhangetalempty}.

The beneficial role of environmental variability with fragmented habitats for tag-based cooperative systems that was identified through simulations with spatially correlated empty sites in our model can also be understood as an indicator of ecosystem antifragility \cite{equigershenson}. It is a novel way of reconceptualizing resilience in complex systems such that structures with antifragile properties are not just resistant to perturbation and other environmental pressures but they can also benefit from perturbation-induced stressful conditions. It will therefore remain a challenge for future research and next generalizations of our current model to further investigate the conditions under which social systems with tag-mediated interactions can both resist to and benefit from elevated environmental stressors such as habitat destruction and the associated resource scarcity. 

One possible way of addressing this question could be by studying the influence of dispersal and migratory behaviors on tag-based cooperation in systems with spatially structured environmental heterogeneities, especially because dispersal can knowingly play a key role in avoiding catastrophic tipping points when habitat loss and fragmentation disrupt metapopulation dynamics \cite{sardenyassole}, and in territorial expansion and successful colonization of new environments under periodic habitat destruction \cite{kanghao}. Here, tags could be used not only as phenotypic features for conditional, tag-mediated cooperative decisions, but also as signals for migratory decisions \cite{dhakal}, e.g. to indicate a movement away from degraded habitats before agents are even able to explore the detailed ecological conditions in the novel environment. 

Future studies should also address co-evolutionary dynamics of tag-based strategies and habitat loss, especially in multiplex networks where habitat clearance can progress at varying rates in different layers of heterogeneous multilayered networked systems. Since human social networks, as well as artificial networked systems, are typically multilayered and characterized by both temporal and spatial clustering of contacts \cite{newmancluster}, it would be interesting to extend our model in the context of coupled dynamics of multiple competing processes \cite{pla} in clustered multiplex networks \cite{wuhadziamm2}, simultaneously capturing e.g. behavioral strategy spreading (in the contact layer) and time-varying diffusion of information \cite{superspreader} about locally degraded habitats (in the information layer). In the contact layer, agents would engage into standard social dilemma based pairwise interactions, and in the information layer, they would spread information about ecologically (un)favorable conditions in the environment, that could further influence their migratory decisions. 

However, in addition to these different research opportunities, we note that future extensions of our work should further address the analysis of environmental heterogeneities in cooperative systems, and their change and coevolution with individual agent behavior, especially in the context of anthropogenic habitat destruction and resource depletion that could in turn lead to the emergence of novel behavioral phenotypes. Intriguingly, such studies could potentially provide us with new ways of thinking about alternative evolutionary pathways that have hitherto remained largely unexplored \cite{hadzibeganovic1,nicheconstruction}. 

In sum, we have seen that pruning the connectivity among agents by leaving certain sites uninhabitable can have widely beneficial effects on the evolution of altruism in phenotypically diverse populations, and that these effects depended non-trivially on the complex interplay between $q_{0|0}$, $r$, and $\lambda$, such that the strongest relationship between $q_{0|0}$ and $\rho_C$ occured when $\lambda$ was high while simultaneously $r$ remained low; likewise, the highest level of altruism was found when $q_{0|0}$ and $r$ were both low while $\lambda$ was sufficiently high. On the other hand, the lowest level of $\rho_C$ was associated with higher values of all three variables $q_{0|0}$, $r$, and $\lambda$. Moreover, higher phenotypic diversity was not able to mitigate the negative effects of elevated $q_{0|0}$ and $r$ on the evolution of altruism in moderately degraded habitats, but it promoted unconditional cooperation if $q_{0|0}$ and $r$ were both sufficiently low. Thus, the effect of $q_{0|0}$ on the density of strategies varied in our model not only by phenotypic diversity $\lambda$, but also by how low or high the cost-to-benefit ratio $r$ was. 

\section{Conclusions}
Habitat fragmentation and loss have long been regarded as the principal threats to species interaction and biodiversity conservation on our planet. Fragmentation of degraded habitats can permanently isolate previously connected populations, causing a reduced gene flow and the associated loss of genetic diversity. Although the assumption that harsh environmental conditions can stimulate cooperation among affected individuals is rather intuitive, the actual consequences of living in degraded and discontinuous habitats for prosocial behavior remain less well understood.

In the present paper, we investigated the effects of heterogeneous fragmentation of unsuitable sites on the evolution of tag-based cooperation in populations with Snowdrift game interactions. Here, we particularly focused on the study of combined effects of spatially correlated empty sites, phenotypic diversity, and varying cost-to-benefit ratio on the evolutionary dynamics of conditional and unconditional strategies. We additionally investigated a simpler, aspatial model, for which we developed analytic arguments describing individual transition probabilities with $K$ tags and four different strategies, evolving in infinite, well-mixed populations under the constraint of weak selection. Our analysis revealed conditions for the two different rankings of unconditional and conditional strategies, showing that the 
fraction of $I$-strategy (ethnocentrism) can dominate under most conditions in well-mixed populations. However, our aspatial model was not able to predict the dominance of pure altruism that was discovered only thanks to extensive agent-based computer simulations of our spatially explicit model of tag-based cooperation with precisely structured environmental heterogeneities.

Different from earlier studies, our spatial model revealed that low to intermediate clustering degree of empty sites in combination with higher phenotypic diversity and low cost-to-benefit ratio $r$ can markedly suppress ethnocentric cooperation while simultaneously promoting pure altruism. However, at intermediate cost-to-benefit ratios such as $r=0.5$, we observed a remarkable advantage of ethnocentric cooperation which outweighed all other competing strategies in all considered model versions, regardless of the number of tags and irrespective of the underlying clustering degree of empty sites. In addition, we found stable coexistence of unconditional and conditional strategies that persisted in a variety of conditions, including a novel case in which strong extra-group cooperation stably coexisted in evolutionary time with ethnocentrism, representing a unique scenario which has rarely been observed in earlier tag-based cooperation models.

Together, these findings highlight a considerable robustness of ethnocentric cooperation in phenotypically less diverse systems with empty sites, but also its striking fragility at lower cost-to-benefit ratios in phenotypically more diverse populations occupying heterogeneously fragmented habitats with scattered arrangement of unsuitable sites. Our findings also demonstrate a remarkable facilitation of pure altruism (i.e. unconditional cooperation) in phenotypically more diverse systems, especially at low $r$ and lower to intermediate clustering degrees of empty sites. Thus, indiscriminate cooperation can flourish and prevail in the presence of multiple phenotypic features and scattered or randomly distributed empty sites in spatially heterogeneous networked systems, even in the presence of other potent competitor strategies such as ethnocentrism. 

Our results further suggest that fragmentation of moderately degraded habitats can function as a cooperation-enhancing mechanism in phenotypically diverse populations, especially when ingroup-biased conditional strategists threaten to eliminate unconditional generosity and take over the ever scarcer resources. These results further indicate that a networked system with tag-based interactions can generally benefit from variably degraded environments, such as those with spatially correlated empty sites, by promoting socially desirable pure altruism and by suppressing potent but discriminating strategies such ethnocentrism.

Our findings could therefore have profound implications for future modeling of ecosystem antifragility in heterogeneous networked populations, as well as for the study of evolution of global cooperation and competition under harsh environmental conditions. These findings may also be instrumental in shaping our further understanding of mechanisms involved in the evolution of ethnocentrism and other potentially spiteful strategies, whose reduction and control could be critical to the prevention of catastrophic tipping points in moderately destructed but fragmented societies.

\section{Acknowledgments}
\label{}
This work was supported by the National Key R\&D Program of China (grant No. 2021YFA1000402) and the National Natural Science Foundation of China (grants Nos. 32271554 and 62073112). 


\appendix

\section{Transition probabilities in the aspatial model}

Using the method described in Section 3, we now calculate other transition probabilities between considered strategies in our aspatial model with $4K$ types of individuals. Here, the transition probability from $C$ to $E$, is given by
\begin{equation}
\rho_{C \rightarrow E}=\frac{\mu}{4K}\left(\frac{1}{N}+\frac{\beta}{6}\left(-b+2c-\frac{b-\frac{c}{2}}{N}\right)\right)+\mu\frac{K-1}{4K}\left(\frac{1}{N}+\frac{\beta}{6}\left(-b+\frac{c}{2}-\frac{b-\frac{c}{2}}{N}\right)\right).
\end{equation}
The transition probability from $C$ to $D$, is then given by
\begin{equation}
\rho_{C \rightarrow D}=\frac{\mu}{4K}\left(\frac{1}{N}+\frac{\beta}{6}\left(-b+2c-\frac{b-\frac{c}{2}}{N}\right)\right)+\mu\frac{K-1}{4K}\left(\frac{1}{N}+\frac{\beta}{6}\left(-b+2c-\frac{b-\frac{c}{2}}{N}\right)\right).
\end{equation}
The transition probability from $I$ to $C$ is 
\begin{equation}
\rho_{I \rightarrow C}=\frac{\mu}{4K}\left(\frac{1}{N}-\frac{\beta}{6}\left(\frac{3b-\frac{3c}{2}}{N}\right)\right)+\mu\frac{K-1}{4K}\left(\frac{1}{N}+\frac{\beta}{6}\left(-\frac{3c}{2}-\frac{3b-\frac{3c}{2}}{N}\right)\right).
\end{equation}
The transition probability from $I$ to $E$ is
\begin{equation}
\rho_{I \rightarrow E}=\frac{\mu}{4K}\left(\frac{1}{N}+\frac{\beta}{6}\left(-b+2c-\frac{b-\frac{c}{2}}{N}\right)\right)+\mu\frac{K-1}{4K}\left(\frac{1}{N}
+\frac{\beta}{6}\left(-b-c-\frac{b-\frac{c}{2}}{N}\right)\right).
\end{equation}
The transition probability from $I$ to $D$ is
\begin{equation}
\rho_{I \rightarrow D}=\frac{\mu}{4K}\left(\frac{1}{N}+\frac{\beta}{6}\left(-b+2c-\frac{b-\frac{c}{2}}{N}\right)\right)+\mu\frac{K-1}{4K}\left(\frac{1}{N}+\frac{\beta}{6}\left(-2b+c-\frac{-b+\frac{c}{2}}{N}\right)\right).
\end{equation}
 
The transition probability from $E$ to $C$ is
\begin{equation}
\rho_{E \rightarrow C}=\frac{\mu}{4K}\left(\frac{1}{N}+\frac{\beta}{6}\left(2b-\frac{5c}{2}-\frac{4b-2c}{N}\right)\right)+\mu\frac{K-1}{4K}\left(\frac{1}{N}
+\frac{\beta}{6}\left(2b-c-\frac{4b-2c}{N}\right)\right).
\end{equation}
The transition probability from $E$ to $I$ is
\begin{equation}
\rho_{E \rightarrow I}=\frac{\mu}{4K}\left(\frac{1}{N}+\frac{\beta}{6}\left(2b-\frac{5c}{2}-\frac{4b-2c}{N}\right)\right)+\mu\frac{K-1}{4K}\left(\frac{1}{N}
+\frac{\beta}{6}\left(2b+\frac{c}{2}-\frac{4b-2c}{N}\right)\right).
\end{equation}
The transition probability from $E$ to $D$ is
\begin{equation}
\rho_{E \rightarrow D}=\frac{\mu}{4KN}+\mu\frac{K-1}{4K}\left(\frac{1}{N}+\frac{\beta}{6}\left(b+c-\frac{2b-c}{N}\right)\right).
\end{equation}
The transition probability from $D$ to $C$ is
\begin{equation}
\rho_{D \rightarrow C}=\frac{\mu}{4K}\left(\frac{1}{N}+\frac{\beta}{6}\left(2b-\frac{5c}{2}-\frac{4b-2c}{N}\right)\right)+\mu\frac{K-1}{4K}\left(\frac{1}{N}
+\frac{\beta}{6}\left(2b-\frac{5c}{2}-\frac{4b-2c}{N}\right)\right),
\end{equation}
and finally, the transition probability from $D$ to $E$ is given by
\begin{equation}
\rho_{D \rightarrow E}=\frac{\mu}{4KN}+\mu\frac{K-1}{4K}\left(\frac{1}{N}+\frac{\beta}{6}\left(b-2c-\frac{2b-c}{N}\right)\right).
\end{equation}

\newpage

\section{Evolutionary dynamics of strategies in 2-tag and 4-tag systems with empty sites}

\begin{figure}[htb]
\begin{center}
\includegraphics[width=15cm]{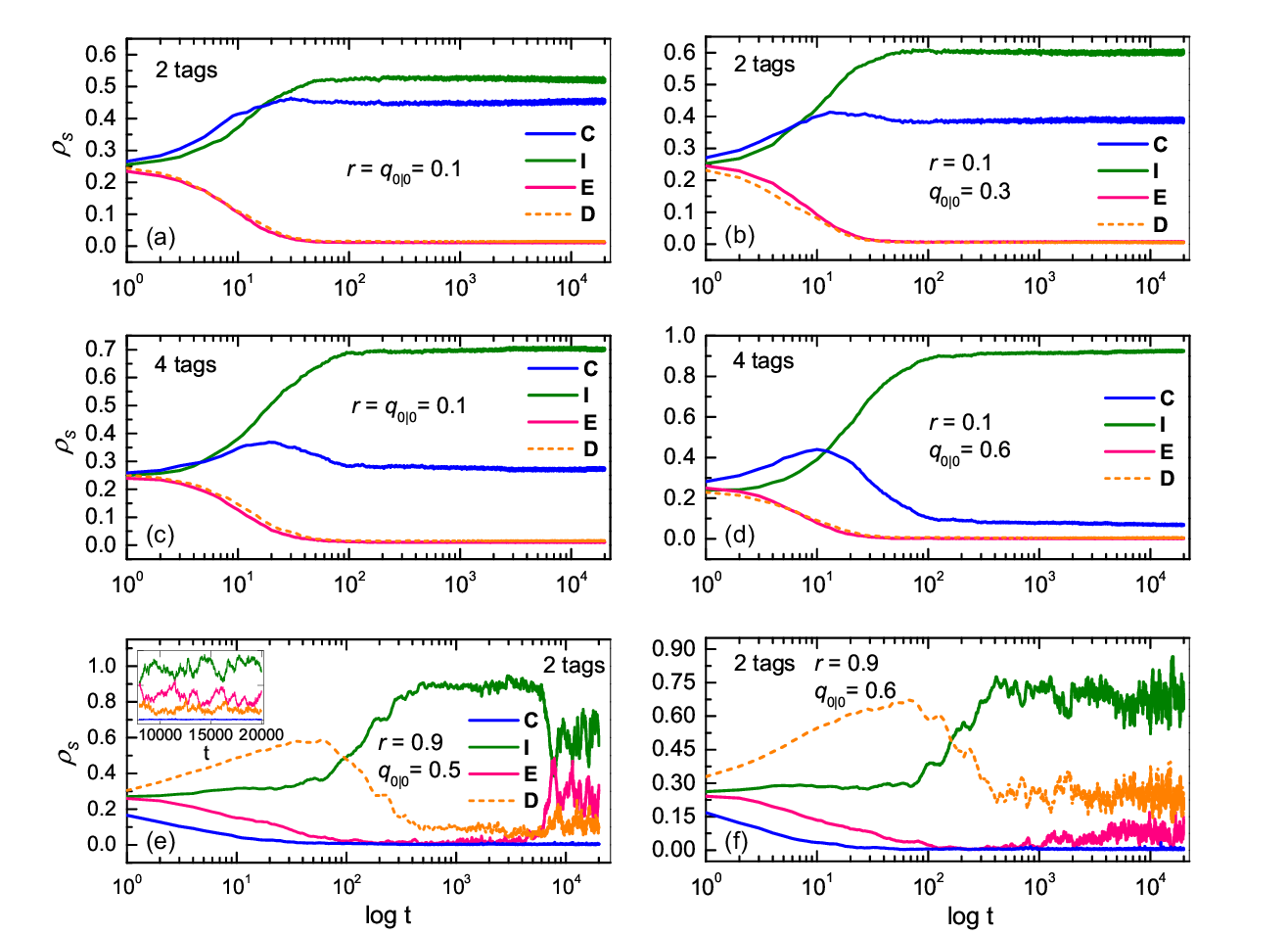}
\caption{Typical time series for the fractions of four strategies in systems with two and four tags, under different cost-to-benefit ratios $r$ and varying clustering degrees of empty sites $q_{0|0}$. The lattice size was $L = 100 \times 100$, and the four strategies (C, I, E, and D) were initially randomly distributed across the habitable lattice sites, with equal frequencies at $t=0$. Panels (a) - (d) illustrate how the difference between the two strongest strategies, I and C, changes as a function of different parameter values. Two conditions yielding oscillatory dynamics of I, E, and D strategies (observed in 2-tag systems only), are shown in panels (e) and (f).}
\end{center}
\end{figure}


\newpage

\begin{thebibliography}{00}

\bibitem{pimmraven} S.L. Pimm, C.N. Jenkins, R. Abell, T.M. Brooks, J.L. Gittleman, L.N. Joppa, P.H. Raven, C.M. Roberts, J.O. Sexton, The biodiversity of species and their rates of extinction, distribution, and protection, Science 344 (2014) 1246752.

\bibitem{lionvanbaalen} S. Lion, M. van Baalen, Self-structuring in spatial evolutionary ecology, Ecol. Lett. 11 (2008) 277--295.
\bibitem{cheptou} P.-O. Cheptou, A.L. Hargreaves, D. Bonte, H. Jacquemyn, Adaptation to fragmentation: Evolutionary dynamics driven by human influences, Phil. Trans. Roy. Soc. B 372 (2017) 20160037.
\bibitem{fahrig} L. Fahrig, Effects of habitat fragmentation on biodiversity, Annu. Rev. Ecol. Evol. Syst. 34 (2003) 487--515.

\bibitem{sardenyassole} J. Sardany\' es, J. Pi\~{n}ero, R. Sol\' e, Habitat loss-induced tipping points in metapopulations with facilitation, Populat. Ecol. 61 (2019) 436--449.
\bibitem{fahrig2} L. Fahrig, Ecological responses to habitat fragmentation per se, Annu. Rev. Ecol. Evol. Syst. 48 (2017) 1--23.
\bibitem{parkermacnally} M. Parker, R. Mac Nally, Habitat loss and the habitat fragmentation threshold: An experimental evaluation of impacts on richness and total abundances using grassland invertebrates, Biol. Conserv. 105 (2002) 217--229.

\bibitem{limafilho2021} J.A. de Lima Filho, R.J. Vieira, C.A.M. de Souza, F.F. Ferreira, V.M. de Oliveira, Effects of habitat fragmentation on 
biodiversity patterns of ecosystems with resource competition, Physica A 564 (2021) 125497.

\bibitem{ZhangBearupetal2023} H. Zhang, D. Bearup, G. Barab\' as, W.F. Fagan, I. Nijs, D. Chen, J. Liao, Complex nonmonotonic responses of biodiversity to 
habitat destruction, Ecology 104 (2023) e4177.

\bibitem{NatEcolEvol2024} H. Zhang, J.M. Chase, J. Liao, Habitat amount modulates biodiversity responses to fragmentation, Nature Ecol. Evolut. 8 (2024) 1437--1447.

\bibitem{zhanghuihan} F. Zhang, C. Hui, X.Z. Han, Z.Z. Li, Evolution of cooperation in patchy habitat under patch decay and isolation, Ecol. Res. 20 (2005) 461--469.

\bibitem{taoetal2024} Y. Tao, A. Hastings, K.D. Lafferty, I. Hanski, O. Ovaskainen, Landscape fragmentation overturns classical metapopulation 
thinking, Proc. Natl. Acad. Sci. USA 121 (2024) e2303846121.

\bibitem{alizon} S. Alizon, P. Taylor, Empty sites can promote altruistic behavior, Evolution 62 (2008) 1335--1344.

\bibitem{sekiguchi} T. Sekiguchi, M. Nakamaru, Effect of the presence of empty sites on the evolution of cooperation by costly punishment in spatial games, J. Theor. Biol. 256 (2008) 297--304.

\bibitem{yangphysa} X. Yang, W. Wang, F. Zhang, H. Qiao, Cooperation enhanced by habitat destruction in Prisoner's Dilemma games, Physica A 486 (2017) 668--673.

\bibitem{yangpla} X. Yang, F. Zhang, W. Wang, D. Zhang, Z. Shi, S. Zhou, Effect of habitat destruction on cooperation in public goods games, 
Phys. Lett. A 384 (2020) 126276.

\bibitem{zhangetalempty} H. Zhang, L. Wang, D.S. Hou, Effect of the spatial autocorrelation of empty sites on the evolution of cooperation, 
Physica A 443 (2016) 296--308.

\bibitem{hamiltonfish} W.D. Hamilton, Geometry for the selfish herd, J. Theor. Biol. 31 (1971) 295--311.
\bibitem{nowakmay} M.A. Nowak, R.M. May, Evolutionary games and spatial chaos, Nature 359 (1992) 826--829.
\bibitem{tarnita} M.A. Nowak, C.E. Tarnita, T. Antal, Evolutionary dynamics in structured populations, Phil. Trans. R. Soc. Lond. 365 (2010) 19--30.
\bibitem{ohtsuki} H. Ohtsuki, C. Hauert, E. Lieberman, M.A. Nowak, A simple rule for the evolution of cooperation on graphs and social networks, 
Nature 441 (2006) 502--505.
\bibitem{randnowak} D.G. Rand, M.A. Nowak, Human cooperation, Trends Cogn. Sci. 17 (2013) 413--425.
\bibitem{fiverules} M.A. Nowak, Five rules for the evolution of cooperation, Science 314 (2006) 1560--1563.

\bibitem{nandadurret} M. Nanda, R. Durret, Spatial evolutionary games with weak selection, Proc. Natl. Acad. Sci. USA 114 (2017) 6046--6051.

\bibitem{masudahetero} N. Masuda, Participation costs dismiss the advantage of heterogeneous networks in evolution of cooperation, Proc. Roy. Soc. B 274 (2007) 1815--1821.


\bibitem{tomassini} M. Tomassini, E. Pestelacci, L. Luthi, Social dilemmas and cooperation in complex networks, Int. J. Mod. Phys. C 18 (2007) 1173--1185.

\bibitem{zhangmathbiol} H. Zhang, A game-theoretical dynamic imitation model on networks, J. Math. Biol. 82 (2021) 1--22.

\bibitem{vanveelen} M. van Veelen, J. Garc\' ia, D.G. Rand, M.A. Nowak, Direct reciprocity in structured populations, Proc. Natl Acad. Sci. USA 109 (2012) 9929--9934. 

\bibitem{huilattice} H. Zhang, M. Gao, Z. Li, Z. Maa, H. Wang, The ambivalent effect of lattice structure on a spatial game, Physica A 390 (2011) 1961--1972.

\bibitem{allenetal} B. Allen, G. Lippner, Y.-T. Chen, B. Fotouhi, M.A. Nowak, S.-T. Yau, Evolutionary dynamics on any population structure, Nature 544 (2017) 227--230.

\bibitem{gracialazaro} C. Gracia-L\' azaro, A. Ferrer, G. Ruiz, A. Taranc\' on, J.A. Cuesta, A. S\' anchez, Y. Moreno, Heterogeneous networks do not promote cooperation when humans play a Prisoner's Dilemma, Proc. Natl. Acad. Sci. USA 109 (2012) 12922--12926. 

\bibitem{cuestasanchez} J.A. Cuesta, C. Gracia-L\' azaro, A. Ferrer, Y. Moreno, A. S\' anchez, Reputation drives cooperative behavior and network formation in human groups, Sci. Rep. 5 (2015) 7843.

\bibitem{grujic} J. Gruji\' c, C. Gracia-L\' azaro, M. Milinski, D. Semmann, A. Traulsen, J.A. Cuesta, Y. Moreno, A. S\' anchez, A comparative analysis of spatial Prisoner's Dilemma: Conditional cooperation and payoff irrelevance, Sci. Rep. 4 (2014) 4615.

\bibitem{mcavoy}A. McAvoy, J. Wakeley, Evaluating the structure-coefficient theorem of evolutionary game theory, Proc. Natl. Acad. Sci. USA 119 (2022) e2119656119.

\bibitem{shixiao} Q. Shi, S. Xiao, Q. Dai, H. Li, J. Yang, Spatial structure might impede cooperation in evolutionary games with reinforcement learning, Int. J. Mod. Phys. C 33 (2022) 2250168.

\bibitem{sinharoy} S. Sinha, D. Nath, S. Roy, Topology dependent payoffs can lead to escape from prisoner's dilemma, Eur. Phys. J. B 94 (2021) 1--7.


\bibitem{svoboda2024} J. Svoboda, K. Chatterjee, Density amplifiers of cooperation for spatial games, Proc. Natl. Acad. Sci. USA 121 (2024) e2405605121.

\bibitem{linli2025} L. Li, J. Lv, J. Ruan, L. Ma, The effects of dynamic peer pressure on the evolution of cooperation in 
complex networks, Physica A 665 (2025) 130489.

\bibitem{holland} J. Holland, The effect of labels (tags) on social interactions, Working Paper 93-10-064, Santa Fe Institute, Sante Fe, New Mexico, 1993.
\bibitem{riolo} R.L. Riolo, M.D. Cohen, R. Axelrod, Evolution of cooperation without reciprocity, Nature 414 (2001) 441--443.
\bibitem{hamaxel} R.A. Hammond, R. Axelrod, The evolution of ethnocentrism, J. Conflict Resolut. 50 (2006) 926-936.
\bibitem{traulsenandnowak} A. Traulsen, M.A. Nowak, Chromodynamics of cooperation in finite populations, PLoS ONE 3 (2007) e270.

\bibitem{antal} T. Antal, H. Ohtsuki, J. Wakeley, P.D. Taylor, M.A. Nowak, Evolution of cooperation by phenotypic similarity, Proc. Natl Acad. Sci. USA 106 (2009) 8597--8600.
\bibitem{laird1} R.A. Laird, Evolutionary strategy dynamics for tag-based cooperation and defection in the spatial and aspatial Snowdrift game, 
Intl. J. Bifurcat. Chaos 22 (2012) 1230039.
\bibitem{garciaveelen} J. Garcia, M. van Veelen, A. Traulsen, Evil green beards: Tag recognition can also be used to withhold cooperation in structured populations, J. Theor. Biol. 360 (2014) 181--186.
\bibitem{hadzibeganovic0} T. Hadzibeganovic, F.W.S. Lima, D. Stauffer, Benefits of memory for the evolution of tag-based cooperation in structured 
populations, Behav. Ecol. Sociobiol. 68 (2014) 1059--1072.
\bibitem{hadzibeganovic1} T. Hadzibeganovic, D. Stauffer, X.P. Han, Randomness in the evolution of cooperation, Behav. Process. 113 (2015) 86--93.

\bibitem{hartshorn} M. Hartshorn, A. Kaznatcheev and T. Shultz, The evolutionary dominance of ethnocentric cooperation, J. Artif. Soc. Soc. Simul. 16 (2013) 7.
\bibitem{hadzibeganovic2} T. Hadzibeganovic, C.Y. Xia, Cooperation and strategy coexistence in a tag-based multi-agent system with contingent mobility, 
Knowl.-Based Syst. 112 (2016) 1--13.
\bibitem{hadzibeganovic3} T. Hadzibeganovic, D. Stauffer, X.P. Han, Interplay between cooperation-enhancing mechanisms in
evolutionary games with tag-mediated interactions, Physica A 496 (2018) 676--690.

\bibitem{ramazi} P. Ramazi, M. Cao, F.J. Weissing, Evolutionary dynamics of homophily and heterophily, Sci. Rep. 6 (2016) 22766.
\bibitem{morsky} B. Morsky, R. Cressman, C.T. Bauch, Homophilic replicator equations, J. Math. Biol. 75 (2017) 309--325.
\bibitem{Jensen2019pre} G.G. Jensen, F. Tischel and S. Bornholdt, Discrimination emerging through spontaneous symmetry breaking in a spatial prisoner's dilemma model with multiple labels, Phys. Rev. E 100 (2019) 062302.
\bibitem{hadzibeganovic4} T. Hadzibeganovic, P.B. Cui, Z.X. Wu, Nonconformity of cooperators promotes the emergence of pure
altruism in tag-based multi-agent networked systems, Knowl.-Based Syst. 171 (2019) 1--24.
\bibitem{hadzibeganovic5} T. Hadzibeganovic, C. Liu, R. Li, Effects of reproductive skew on the evolution of ethnocentrism in structured populations with variable size, Physica A 568 (2021) 125550.
\bibitem{dhakal} S. Dhakal, R. Chiong, M. Chica, T.A. Han, Evolution of cooperation and trust in an N-player social dilemma game with tags for migration decisions, Roy. Soc. Open Sci. 9 (2022) 212000.
\bibitem{jeongetal} W. Jeong, T. Hadzibeganovic, U. Yu, Evolution of cooperation with time-varying tags and heterogeneous immigration dynamics, Int. J. Mod. Phys. C 33 (2022) 2350010.
\bibitem{HeLiDu2025} X. He, G. Li, H. Du, Effects of tag mediation and structural balance on the evolution of cooperation on signed networks, Chaos 35 (2025) 043113.
\bibitem{HeChengLu2025} X. He, G. Cheng, J. Lu, Tag-mediated effect on the dynamics of social influence, PLoS One 20 (2025) e0338598.

\bibitem{efferson} C. Efferson, R. Lalive and E. Fehr, The coevolution of cultural groups and ingroup favoritism, Science 321 (2008) 1844--1849.

\bibitem{masufu} N. Masuda, F. Fu, Evolutionary models of in-group favoritism, F1000Prime Rep. 7 (2015) 27.

\bibitem{santospalattice} F.C. Santos, F.L. Pinheiro, T. Lenaerts, J.M. Pacheco, The role of diversity in the evolution of cooperation, J. Theor. Biol. 299 (2012) 88--96.

\bibitem{galliard} J.-F. Le Galliard, R. Ferri\'{e}re, U. Dieckmann, The adaptive dynamics of altruism in spatially heterogeneous populations, Evolution 57 (2003) 1--17.

\bibitem{hiebeler1} D. Hiebeler, Populations on fragmented landscapes with spatially structured heterogeneities: Landscape generation and local dispersal, Ecology 81 (2000) 1629--1641.

\bibitem{nonacs} P. Nonacs, Reproductive skew in cooperative breeding: Environmental variability, antagonistic selection, choice, and control, Ecol. Evol. 9 (2019) 10163--10175.

\bibitem{janssonjtb} F. Jansson, What games support the evolution of an ingroup bias? J. Theor. Biol. 373 (2015) 100--110.

\bibitem{barreira} A. Barreira da Silva Rocha, A. Laruelle, Evolution of cooperation in the snowdrift game with heterogeneous population, 
Adv. Complex Syst. 16 (2013) 1350036.

\bibitem{greenwood} G.W. Greenwood, Enhanced cooperation in the N-person iterated snowdrift game through tag mediation, 2011 IEEE Conference on Computational Intelligence and Games (CIG'11), pp. 1--8, 2011.

\bibitem{pdsd} F. Fu, M.A. Nowak, C. Hauert, Invasion and expansion of cooperators in lattice populations: Prisoner's dilemma vs. snowdrift games, J. Theor. Biol. 266 (2010) 358--366. 

\bibitem{haudo2004} C. Hauert, M. Doebeli, Spatial structure often inhibits the evolution of cooperation in the Snowdrift game, Nature 428 (2004) 643--646.

\bibitem{tanimotofactorial} A. Yamauchi, J. Tanimoto, A. Hagishima, An analysis of network reciprocity in Prisoner's Dilemma games using 
Full Factorial Designs of Experiment, BioSystems 103 (2011) 85--92.

\bibitem{taniki} J. Tanimoto, N. Kishimoto, Network reciprocity created in prisoner's dilemma games by coupling two 
mechanisms, Phys. Rev. E 91 (2015) 042106.


\bibitem{szabofath2007} G. Szab\'{o}, G. F\' ath, Evolutionary games on graphs, Phys. Rep. 446 (2007) 97--216.


\bibitem{adamiabm} C. Adami, J. Schossau, A. Hintze, Evolutionary game theory using agent-based methods, Phys. Life Rev. 19 (2016) 1--26.

\bibitem{axelhamil} R. Axelrod, W.D. Hamilton, The evolution of cooperation,  Science 211 (1981) 1390--1396.

\bibitem{maynard} J. Maynard Smith, Evolution and the Theory of Games, Cambridge University Press, Cambridge, UK, 1982.
\bibitem{doebelihau} M. Doebeli, C. Hauert, Models of cooperation based on the Prisoner's Dilemma and the Snowdrift game, Ecol. Lett. 8 (2005) 748--766.


\bibitem{hofsig98} J. Hofbauer, K. Sigmund, Evolutionary Games and Population Dynamics, Cambridge University Press, Cambridge, 1998.

\bibitem{nowak2006} M.A. Nowak, Evolutionary Dynamics: Exploring the Equations of Life, Harvard University Press, Cambridge,  2006.

\bibitem{traulsen2006} A. Traulsen, M.A. Nowak, J.M. Pacheco, Stochastic dynamics of invasion and fixation, Phys. Rev. E 74 (2006) 11909.

\bibitem{szabofermi} G. Szab\'{o}, C. T\H{o}ke, Evolutionary prisoner's dilemma game on a square lattice, Phys. Rev. E 58 (1998) 69--73.

\bibitem{traulsen371} A. Traulsen, Mechanisms for similarity based cooperation, Eur. Phys. J. B 63 (2008) 363--371.


\bibitem{hauertetal2007} C. Hauert, A. Traulsen, H. Brandt, M.A. Nowak, K. Sigmund, Science 316 (2007) 1905.

\bibitem{satoiwasa} K. Sat$\bar{o}$, Y. Iwasa, Pair approximations for lattice-based ecological models, In: U. Dieckmann, R. Law, J.A.J. Metz (Eds.), The geometry of ecological Interactions: simplifying spatial complexity, Cambridge University Press, pp.341-358, 2000. 

\bibitem{huietal2006} C. Hui, M.A. MeGeoch, M. Warren, A spatially explicit approach to estimating species occupancy and spatial correlation, J. Anim. Ecol. 75 (2006) 140--147.

\bibitem{hiebeler2} D.E. Hiebeler, B.R. Morin, The effect of static and dynamic spatially structured disturbances on a locally dispersing population, J. Theor. Biol. 246 (2007) 136--144.


\bibitem{hauertholmes} C. Hauert, M. Holmes, M. Doebeli, Evolutionary games and population dynamics: Maintenance of cooperation in public goods games, 
Proc. R. Soc. B 273 (2006) 2565--2570 .

\bibitem{masudajtb} N. Masuda, Ingroup favoritism and intergroup cooperation under indirect reciprocity based on group reputation, J. Theor. Biol. 311 (2012) 8--18.

\bibitem{mourishultz} C.J. Mouri, T.R. Shultz, Cooperative intergroup mating can overcome ethnocentrism in diverse populations, arXiv:1504.08312v1, 2015.

\bibitem{grossetal2023} J. Gross, Z.Z. Meder, C.K.W. De Dreu, A. Romano, W.E. Molenmaker, L.C. Hoenig, The evolution of universal cooperation, 
Sci. Adv. 9 (2023) eadd8289.

\bibitem{equigershenson} M. Equihua, M. Espinosa Aldama, C. Gershenson, O. Lopez-Corona, M. Munguia, O. Perez-Maqueo, E. Ramirez-Carrillo, Ecosystem antifragility: Beyond integrity and resilience, PeerJ 8 (2020) e8533. 
 
\bibitem{kanghao} Z.-X. Tan, K.H. Cheong, Periodic habitat destruction and migration can paradoxically enable sustainable territorial expansion, Nonlin. Dyn. 98 (2019) 1--13.

\bibitem{newmancluster} M.E.J. Newman, Random graphs with clustering, Phys. Rev. Lett. 103 (5) (2009) 058701.

\bibitem{pla} Q.C. Wu, T. Hadzibeganovic, X.-P. Han, Coupled dynamics of endemic disease transmission and gradual awareness diffusion in 
multiplex networks, Math. Mod. Meth. Appl. Sci. 33 (2023) 2785--2821.


\bibitem{wuhadziamm2} Q.C. Wu, T. Hadzibeganovic, An individual-based modeling framework for infectious disease spreading in clustered complex networks, Appl. Math. Model. 83 (2020) 1--12.

\bibitem{superspreader} W. Su, X. Wang, G. Chen, Y. Yu, T. Hadzibeganovic, Noise-based synchronization of bounded confidence opinion dynamics in heterogeneous time-varying communication networks, Inf. Sci. 528 (2020) 219--230.

\bibitem{nicheconstruction} G. Gottlieb, Developmental-behavioral initiation of evolutionary change, Psychol. Rev. 109 (2002) 211--218.

\end{thebibliography}
\end{document}